\RequirePackage{fix-cm}
\documentclass[smallextended]{svjour3}       
\smartqed  
\usepackage{booktabs} 

\usepackage[ruled]{algorithm2e} 
\usepackage{array} 
\usepackage{wrapfig}
\usepackage{graphicx}
\usepackage{longtable}
\usepackage{lineno}
\usepackage{framed}
\usepackage{float}
\restylefloat{table}
\usepackage{rotating}
\usepackage{subfigure}
\usepackage{enumitem}
\usepackage{color}
\usepackage{bbding}

\usepackage{wrapfig}
\usepackage{longtable}
\usepackage{lineno}
\usepackage{framed}
\usepackage{float}
\restylefloat{table}
\usepackage{rotating}
\usepackage{enumitem}
\usepackage{color}
\usepackage{framed}
\usepackage{colortbl}
\usepackage{amsmath}

\usepackage{tabularx}
\usepackage{url}
\usepackage{xcolor}
\usepackage{bbding}
\usepackage{url}
\usepackage{balance}
\usepackage{multirow}
\usepackage{booktabs}
\usepackage{graphicx}
\usepackage[framemethod=tikz]{mdframed}

\usepackage{amsmath}
\usepackage{amssymb}
\usepackage{graphicx}
\usepackage{rotating}

\usepackage[ruled]{algorithm2e}

\usepackage{array} 
\usepackage{wrapfig}
\usepackage{graphicx}
\usepackage{longtable}
\usepackage{lineno}
\usepackage{framed}
\usepackage{float}
\usepackage{rotating}
\usepackage{subfigure}
\usepackage{enumitem}
\usepackage{color}
\usepackage{framed}
\usepackage{colortbl}
\usepackage{amsmath}
\usepackage{subfigure}
\usepackage{url}
\usepackage{xcolor}
\usepackage{bbding}
\usepackage{url}

\usepackage{amsmath}

\usepackage{algorithmic}

\newmdenv[innerlinewidth=0.5pt, roundcorner=4pt,innerleftmargin=6pt,
innerrightmargin=6pt,innertopmargin=6pt,innerbottommargin=6pt]{mybox}

\usepackage{array}
\usepackage{multirow}
\usepackage{hhline}

\begin{document}


\title{Organisational Structure Patterns in Agile Teams: An Industrial Empirical Study}


\author{Damian A. Tamburri \and
        Rick Kazman \and 
        Hamed Fahimi
}

\institute{TU/e - JADS \at
              s'Hertogenbosch, The Netherlands \\
              \email{d.a.tamburri@tue.nl}  
                                   \and
              ~SEI/CMU \& University of Hawaii \at
              Honolulu, Hawaii \\
              \email{kazman@hawaii.edu}
               \and
              ~CGI Corp. \at
              Amsterdam, The Netherlands \\
              \email{h.fahimi@cgi.nl}
}

\hyphenation{op-tical net-works semi-conduc-tor}

\maketitle

\begin{abstract}

Forming members of an organization into coherent groups or communities is an important issue in any large-scale software engineering endeavour, especially so in agile software development teams which rely heavily on self-organisation and organisational flexibility.  To address this problem, many researchers and practitioners have advocated a strategy of mirroring system structure and organisational structure, to simplify communication and coordination of collaborative work. But what are the patterns of organisation found in practice in agile software communities and how effective are those patterns?  We address these research questions using mixed-methods research in industry---that is, interview surveys, focus-groups, and delphi studies of agile teams.  In our study of 30 agile software organisations we found that, out of 7 organisational structure patterns that recur across our dataset, a single organisational pattern occurs over 37\% of the time. This pattern: (a) reflects young communities (1-12 months old); (b) disappears in established ones (13+ months); (c) reflects the highest number of architecture issues reported. Finally, we observe a negative correlation between a proposed organisational measure and architecture issues. These insights may serve to aid architects in designing not only their architectures but also their communities to best support their co-evolution.

\end{abstract}
 



\section{Introduction}\label{sec:introduction}

It is widely accepted that organising and shepherding software engineers into healthy communities is important to the success of large-scale agile software development projects \cite{glass,archcomm,Jassowski12}.  But are there \emph{recurrent organisational structural patterns} that can be used as guides for how to create high-functioning agile communities? Informally, an organisational structure pattern is a set of organisational characteristics (e.g., experience diversity or task-allocation policy type \cite{Faludi2006}) and socio-technical characteristics (e.g., socio-technical congruence \cite{congruence}) which are directly measurable from a social network of software practitioners \cite{sna,ossslr}.  Several studies have analysed organisational structures, most notably the seminal paper by Nagappan et al. \cite{nagappan08}. And several studies have concentrated on establishing and evaluating the relationships between the \emph{organisational structure} and the \emph{qualities} of software, but have not identified the organisational structure patterns and characteristics \cite{ossslr} that may affect agile software development projects and organisational structures. Conversely, in related disciplines such as cognitive ergonomics, organisational culture, organizations research, operations research, as well as social networks analysis there are entire theories and knowledge frameworks around governing and efficiently steering data-driven organisational structures.

To fill this gap, we offer an exploratory industrial study seeking to understand the relationships between organisational structures and their efficacy---using as a proxy for such efficacy the software structures being maintained by those structures---with a goal of offering practical insights into such structures and their operational parameters. This is critical for several reasons. First, much literature \cite{KwanCD12,herbsleb1999beyondconwayslaw,paperaranda2008,bh98} and debate \cite{HerbslebG99,SyeedH13} has focused on this  mapping without clearly providing rules, methods, and guidelines that practitioners can use to change or influence the parameters in their own organizations. Second, to enable the inception of data-driven software organizations\footnote{\url{https://www.cio.com/article/3449117/what-exactly-is-a-data-driven-organization.html}}, the future of software development and operations beyond DevOps \cite{Bass17}. 

Our research objective is to find evidence, in agile teams, of a bijective relationship between software organizations and software designs, as was noted by Colfer and Baldwin in their ``Mirroring Hypothesis'' \cite{colfer2010mirroring}. The negative consequences of misalignments has been explored (mostly qualitatively) in software architecture research \cite{MartiniBC14,DingLTV15,MoCKX15}. 
Specifically we seek to understand the ramifications (if any) of social structures on the health of software architectures.

In this article, we therefore address two research questions: (1) \emph{is there a recurrent organisational structure pattern in agile software engineering teams?}; (2) \emph{if so, what does that pattern imply, in terms of software architecture quality?}  Specifically, our research objectives are: (1) to find recurrent organisational structure patterns in modern software engineering projects, using organisational types known in organisational and social-networks research, e.g., Communities of Practice \cite{ossslr}; (2) to find and elaborate the relationships between such patterns and the projects' software architectures, understanding if and how these relationships evolve over time, and their consequences.

For the above purposes, through industrial mixed-methods research, we elicited: (a) organisational structure types and characteristics, i.e., a project's \emph{communication structures}; (b) architectural structure types and characteristics; (c) the issues or problems observed by practitioners in (a) or (b). 

For (a) we used state-of-the-art models, patterns, and characteristics \cite{ossslr} as well as qualitative community typing mechanisms, organisational type identifiers and a type decision-tree that we inferred and validated in our prior work \cite{icgseoss,specissue}. For (b) we used the architecture patterns and characteristics available from the Microsoft Application Architecture Guide\footnote{\url{https://msdn.microsoft.com/en-us/library/ff650706.aspx}}, a widespread and practitioner-focused software architecture reference.  For (c) we used the analogies of software architecture smells \cite{Moha2010,Bertran11,MoCKX15} as well as community smells and analyses from our prior work\footnote{In the scope of this paper we concentrate on reporting architecture issues as a novel contribution, while a detailed overview of community smells and other data obtained in the context of this study is available in \cite{archcomm}} \cite{archcomm,jisaspecissue}. 

Our dataset draws from 30 agile software development organizations in 9 large companies who work with established agile methods  (e.g., Scrum \cite{SchwaberBeedle02}). We chose to focus on organizations that claimed to have recently adopted such methods since agile adoption is reportedly difficult and causes heavy organisational strains \cite{Baham16,IivariI10}.  Our assumption was that \emph{if} there are in fact recurrent and  effective organisational structures in software engineering, we would be able to find them in organisations that recently underwent significant ``rewiring''. 

Our conjectures were that: (a) there exist recurrent (anti)-patterns in the structure of the organizations; and (b) if such organisational structure patterns can be found, then there is an increase in organisational efficiency between ``young" and ``veteran" teams' patterns, according to the principle of organisational stability \cite{QuintanePRM13,WangK01}.

We found 7 organisational structure patterns that occur more than once across the 30 projects in our sample; for example, the ``Formally-Networked, Distributed Informal Communities" Pattern where a formally-structured network (such as the Apache Software Foundation) unites several informal communities (for example, collocated open-source Apache projects). However, one of the reported patterns was found in over 37\% of our dataset: the ``Informally-Networked, Situated, Informal Working Groups" pattern, which reflects informal networks of collocated, tightly-knit, cohesive, and long-lived groups, i.e., teams that do not disband after a single project is over.  This pattern is  reminiscent of the Scrum-of-scrums way of working, where informal networks of tightly-knit and cohesive teams share a non-situated practice. We also observed that the above pattern has 2 \emph{variants}, which differ by a single organisational characteristic, namely: 1) lack of situated action, and 2) lack of organisational goal. 

Contrary to our initial conjectures, the most frequently recurring pattern \textbf{does not} reflect ``veteran" \textbf{nor} high-quality software architectures;  rather, \textbf{the exact opposite} holds true. This pattern appears in young communities and disappears in older communities, whose organisational structure characteristics diverge towards \emph{organisational formality} and \emph{informality} almost equally. In addition, to quantitatively assess the relation behind these observations, we elaborated a simple organisational measure of our own design, i.e., a count of organisational characteristics that a project team clearly exhibits.  Evaluating this characterisation index against our dataset, we observed a moderate negative correlation with reported architecture issues, meaning that an increase of organisational characterisation seems to lead to higher quality architectures. 

From the above findings we conclude that a symbiotic relation  exists between software architectures and the organisational structures associated with them. However, this relation is by no means \emph{positive} and may well be changing continuously. Further research is needed into this relation to quantitatively assess the involved variables.

This article offers four novel contributions to practitioners and researchers: (a) a single empirically-elicited organisational structure pattern reflecting over 1/3 of the agile software organizations we studied---this pattern deserves further attention for organisational quantification and quality measurement; (b) a systematic investigation of which key organisational characteristics are relevant for software organizations---these characteristics can be used by practitioners to steer their organisational ``rewiring" exercises focusing around two organisational characteristics, namely, \emph{team cohesion}, and \emph{team culture}; (c) empirical evidence of the implications of organisational mirroring from industrial practice; (d) an organisational measure and the evaluation of its role as a predictor of software architecture qualities---the measurement can assist practitioners as a means to track the quality of agile software architectures and organizations.

Practitioners may use the insights in this article as a compass to guide their own organisational structures (for example, while migrating to agile methods) and software architectures (modularizing architectures to better suit a division of work among teams), all the while monitoring and maintaining their co-evolution.

\textbf{Structure of the paper.} Section \ref{bg} briefly outlines the background of this study, providing terms and definitions as well. Section \ref{rd} outlines our research design. Later, Section \ref{res} shows our results while Section \ref{disc} discusses our contributions and their threats to validity. Section \ref{rel} outlines related work. Sections \ref{sota} and \ref{conc} place our work in context and offer concluding remarks hinting to future work.

\section{Background and Scope of Our Study}\label{bg}

This section offers an overview of the main research objectives of this article, namely organisational structures (see Sec. \ref{struct}) and software architectures (see Sec. \ref{arch}), providing definitions and conceptual overviews essential for understanding and scoping our study.

\subsection{Organisational Structures}\label{struct}

The literature in organisational structure research primarily focuses on the following areas:
\begin{itemize}
\item Organisational design research - in this field organisational structure types and characteristics are analyzed in terms of their consequences for \emph{organisational design}, i.e., the activity of planning a strategic agenda around a specified organisational structure \cite{Chatha2003};
\item Social-Network Analysis - in this field organisational structure types and characteristics are measurable quantities that can augment social-networks from any context or domain (networks of people, communities of partners, networks of organisations);
\item Cognitive Ergonomics - in this field organisational structure types represent models that allow reasoning on transactive-memory processes \cite{NevoW05}, information representation, as well as information exchange policies;
\end{itemize}

In our prior work \cite{ossslr} we summarised the insights on organisational structures from these fields into common themes or \emph{types} of structures. A structure type is represented by a set of measurable or otherwise evident organisational characteristics (e.g., the presence of informal communication channels across an organisation). For example, an Informal Network is a structure type which characterises a social network where *all* interactions are *always* informal. The way of working within the structure may change radically based on how type characteristics influence that structure. For example, the way of working in a Community of Practice (collocated, tightly knit, practice-focused) is very different than that of a Formal Network (formal, distributed, protocol-based). In particular, According to Mohr \cite{GVK027241513} and later to Lim, Griffiths, and Sambrook \cite{sambrook}, an organisation is a fluid social network where ``organizational structure development is very much dependent on the expression of the strategies and behavior of the [members]" - in an attempt to operationalise, albeit qualitatively, such expression of strategies and behaviour, we look at the expression level of the most observable characteristics in the organisations targeted by our study. More precisely:

\begin{center}
\emph{Organisational Structure Type:\\}
$\omega = \sum \delta(C_1)...\delta(C_n)$;
\end{center}

Where $\omega$ represents the organisational structure type as a sum of characteristics (C$_1-n$) quantified by means of an observability function $\delta$, i.e., a function which assigns a Likert-scale value based on the level of influence that each characteristic bears on the structure. The sum we employ above, is to be interpreted as ``the combined effect of" the most observable characteristics per each type existing across the social network emerging across the structure---the choice of the sum operator, is justified by the fact (established from literature) that the combination of factors affecting the organisational-social network is indeed linear (hence, the sum operator) because of the characteristics? nature of unique identifiers for structure types. For example, an Informal Network type, is strongly indicative of \emph{informal communications} and likely leading to \emph{engaging members}, characterised as follows:

\begin{center}
\emph{Organisational Structure Type IN:\\}
$IN = [\text{\emph{Informality(High),Engagement(Low)}}]$;
\end{center}

Manifestations of these characteristics are typically measured using the average perception of community members over those characteristics \cite{Prandy2000,Mislove2007,Ryynanen12}. For example, people in a community may perceive formality as high or cohesiveness as low, and so on. From this definition, it becomes evident that different types may have a high or low manifestation of the same characteristic; if characteristic X has its highest manifestation in a certain type, X becomes an \emph{identifier}, that is, a primary characteristic for that type and necessary for its identification \cite{ossslr}. For example, Formality is a primary identifier for organisational structures with well defined rules and regulations, typically dictated by corporate governance. Conversely, some types may prescribe the way in which the value of the community is brought about. For example, Informality or Engagement are primary structural identifiers in structures that rely on informal communication, e.g., open-source. Moreover, other types may specify the restrictions that constrain all members of the community, rather than specifying \emph{how} the community operates. For example, High-cohesion and duration are primary in small team scenarios, where 3-10 people are bound to a timeframe by contractual arrangements that establish a small and cohesive project team. 

In the research literature we find evidence of 13 general organisational structure types \cite{ossslr}. Table \ref{overview} offers an overview of such types with a brief introduction to their main characteristics.\footnote{Detailed information and full characterisation for each type is available in our previous work \cite{ossslr}} Furthermore, Figure \ref{outline} (tailored from \cite{archcomm}) offers an overview of types mapped to their primary \emph{identifiers} \cite{ossslr,specissue} - the characteristics that permit their identification. 

\begin{figure*}
\begin{center}
\includegraphics[scale=0.55]{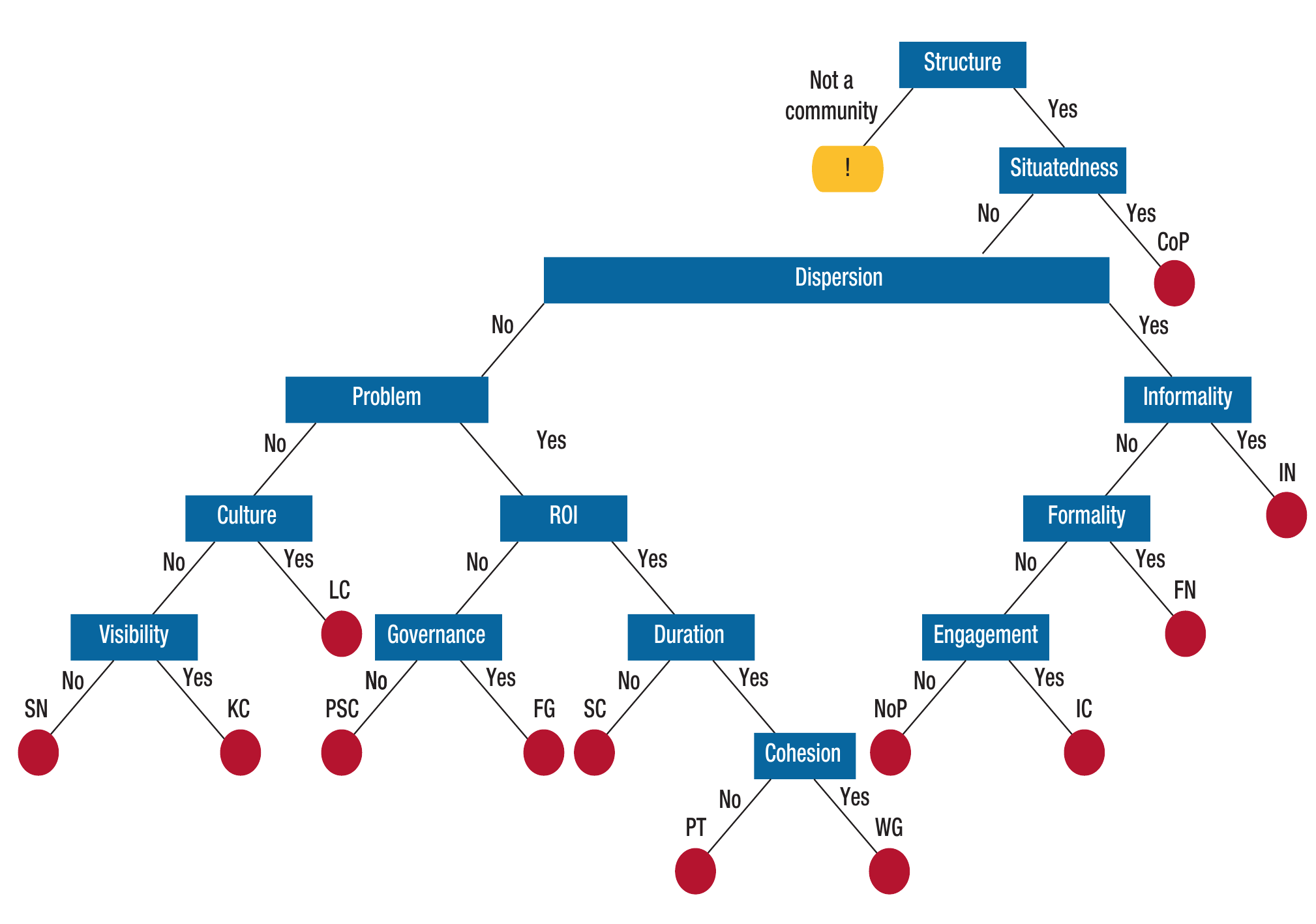}
\end{center}
\caption{A Decision-tree for Organisational Structures, tailored from  \cite{specissue}.}\label{tree}
\end{figure*}

In the scope of this study, and confirming previous results \cite{specissue,icgseoss}, we expect that each data-point in our sample (i.e., each project) exhibits a \emph{set} of  organisational types. This is because the teams and communities in our sample may reflect characteristics of one type in one lifecycle phase, and characteristics of a different type in other phases. Therefore we define:

\begin{center}
\emph{Organisational Structure Pattern:\\}
$\phi = \sum \alpha(t_1)...\alpha(t_n)$;
\end{center}

where t$_{1...n}$ represent the organisational structure types, while $\alpha$ represents the lifecycle-phase in which that type is prominent. While we do not exhaustively elaborate all types and their characteristics in either Fig. \ref{tree} or Table \ref{overview}, the results Section (see Sec. \ref{res}) elaborates the organisational structure types and characteristics which appear in the recurrent patterns found as part of this study. 

Finally, in the scope of this study we define an \emph{organisational characterisation index} as the sum of characteristics evident and measurable for the organisational structure pattern, regardless of the types, namely:

\begin{center}
\emph{organisational characterisation index:\\}
$\omega_m = \sum C_i$;
\end{center}

where $\omega_m$ is the measurement index for organisational structure type $\omega$ while C$_i$ is the set of characteristics (e.g., selected from Fig. \ref{tree} or from literature \cite{ossslr}) reflecting that organisational structure type. 

We are interested to determine whether this measure reflects a positive or negative correlation with respect to reported software architecture issues. The hypothesis we will evaluate is, as introduced in Sec. \ref{sec:introduction}, whether a more complex, governed, and formal organisational structure results in better architectures. Specifically, an increased number of organisational characteristics would reflect a higher organisational measure, and we hypothesize that such organizations would report a lower number of architecture issues (see Sec. \ref{arch}).

\subsection{Organisational Characteristics Explained}\label{sec:characteristics}

As previously stated, the literature reports on as many as 90 characteristics that may manifest organisational structures \cite{ossslr}. This section outlines the characteristics (see Tables \ref{overview} and \ref{outline}) we focus on as primary identifiers of organisational structure types. More specifically, Table \ref{overview} reports en explanation of every single organisational type (column 1 ID and colummn 2 ID expansion) with a brief description (column 3) while Table \ref{outline} details which characteristic corresponds to identifying which type. In addition, a description of every characteristic is reported below.

\begin{itemize}
\item \textbf{Structure.} The observed set of people must be in an organised form and this form must exhibit a structure, i.e. the ``whole" can be split into its ``parts". In a social network, it is always possible to distinguish between the overall structure (macro-structure) and single (sets of) people (micro-structures). A trivial structure is, for example, an unconnected network (i.e. a network of similar but unrelated nodes).
\item \textbf{Situatedness.} If situatedness is prominent in the organisational structure, then people share a common practice using a number of physically co-located social relations and interactions. These interactions lead the members to help each other (hence the term Community) in fostering, maintaining and improving their situated practice \cite{situatedness}.  ``Being situated in this sense is different from simply being located someplace in the way a non-living, non-experiencing object is located. That the body is always situated involves certain kinds of physical and social interactions, and it means that experience is always both physically and socially situated" \cite{situatedness}.
\item \textbf{Dispersion.} If dispersion is prominent in the organisational structure then people share a common practice in a highly-dispersed manner which is exclusively supported by means of digital technologies. Organisational bodies in this condition are often if not always separated by physical, time, and cultural distances that need to be explicitly addressed.
\item \textbf{Informality.} If informality is prominent in the organisational structure, then people are able to communicate freely, through informal interactions (collocated or otherwise). They are permitted to exchange knowledge and artefacts without any protocol or restriction across any kind of distance.
\item \textbf{Duration.} If duration is prominent in the organisational structure, then responsible organisations around the structure have formally determined the longevity of the structure, e.g., based on pre-specified project deadlines. For example, in a software project team, a team may disband upon completion of the project.
\item \textbf{Visibility-Tracking.} If visibility-tracking is prominent in the organisational structure, then the motivation for members of the organisational structure is to capture, cater for, and disseminate knowledge and artefacts; the goal of the structure is making artefacts visible and readily available to all who might be interested.
\item \textbf{Cohesion.} If cohesion is prominent across the organisational structure then the members jointly focus on a particular area of interest in a tightly-knit social, socio-technical, and organisational fashion. A parallel of high-cohesion in software engineering organisational structures is pair-program\-ming, which has been reported as radically changing the software organisational structure \cite{XPImpact}.
\item \textbf{ROI-tracking.} If the tracking of Return-On-Investment is prominent across the organisational structure, then the structure includes very experienced professionals that are contracted by an external sponsor to evaluate and demonstrate the ROI of the  organisational sponsor actions. A parallel in software engineering would be a major software-house (the sponsor) which contracts a consulting firm (the ROI-tracking structure) to double-check and confirm the value of the architecture decisions taken across the software house.
\item \textbf{Engagement.} If engagement is prominent across the organisational structure, then the success of the structure and its goals depends on the degree to which members are actively engaged in it. A parallel in software engineering would be a software Small-Medium Enterprise (SME) whose members are highly-engaged and cater for the success of the company.
\item \textbf{Culture-Tracking.} If culture-tracking mechanisms are prominent across the organisational structure then members are skilled professionals who disclose their professional experience, skills, and problem-solving best-practices for the benefit of all members in the community. A parallel in software engineering would be the StackOverflow learning community.
\item \textbf{Organisational Rigour.} If formality is prominent across the organisational structure, then members are formally appointed and regulated by organisational sponsors; each member's role is to coordinate the structure's resources for the benefit of the entire organisation. These conditions amount to an increased level of organisational rigour. Candidate members are carefully evaluated and selected before membership is granted. Status is formally approved after the formal evaluation. This process guarantees the compatibility of members with the structure goals. A parallel in software engineering would be the Object Management Group (OMG), a standardisation body where members need to participate with considerable financial, human-, and knowledge-capital expenses.
\item \textbf{Problem-Focus.} If problem-focus is prominent across the structure, then structure members are highly-skilled, senior experts who are formally selected and approved by corporate sponsors. These experts brainstorm solutions to problems (organisational, technical, social, or otherwise). Experts are expected to procure a best-practice that solves the problem given a  context.
\item \textbf{Governance.} If governance is prominent across the structure, then structure members need to adhere to formal governance guidelines (software governance \cite{Abbas09}, in our case). Governance guidelines make sure that members comply with specified organisational policies, follow adopted standards, and comply with regulations. A parallel in software engineering reflects organisations with CMMI level certifications to which they adhere.
\end{itemize}

\begin{table*}
\caption{Organisational Structure Types, tailored from \cite{ossslr}.}\label{overview}
\scriptsize

\begin{tabular}{|>{\raggedright}p{.5cm}|>{\raggedright}p{1.5cm}|>{\raggedright}p{9cm}|}
\hline 
\textbf{Label} & \textbf{Name} & \textbf{Description}\tabularnewline
\hline 
COP & Communities of practice & A CoP consists of collocated groups of people who share a concern, a set of problems or a practice. These people interact frequently, face-to-face, collaboratively (mutuality to help each other) --- this set of social processes is called situatedness \cite{situated,Ruikar2009}. For example, software architecture board in a company is a collocated organisational body (i.e., a community) specialised in collaboratively addressing software architecture problems (i.e., the practice).\tabularnewline
\hline 
IN & Informal Networks & INs are loose networks of ties between geographically distributed individuals that
happen to come in contact in a context. Strength of informal ties is the critical bond keeping the community together. INs differ from other types since it does not use governance
practices \cite{Cross2005}. Many free-lance open-source communities are Informal Networks. \tabularnewline
\hline 
FN & Formal Networks & Within FNs, members are rigorously selected, appointed, and governed. They are defined and acknowledged by a management body. Direction is carried out according to defined governance schemes and corporate strategy \cite{ossslr}. Globally-distributed business divisions in high-tech firms often develop into Formal Networks. \tabularnewline
\hline 
IC & Informal Communities & ICs are usually sets of people in an organisation with a common
interest, often closely dependent on their practice. They interact informally, usually across a distance, frequently based on a common history or culture (e.g. shared ideas, experience etc). The main difference they have with all communities (with the exception of NoPs) is that they are necessarily dispersed so that the community can reach a wider audience \cite{ossslr}. StackOverflow is a very good example of an Informal Community around software engineering.\tabularnewline
\hline 
NOP & Networks of Practice & A NoP is a networked system of communication and collaboration that
connects CoPs (which are localized). In principle anyone can join without selection of candidates (e.g. OpenSource forges are instances). NoPs have a high geodispersion, i.e. they can
span geographical and time distances this increases visibility and reachability of members. An unspoken
requirement for entry is IT literacy \cite{Ruikar2009}. A distributed standardisation group such as OASIS\footnote{\url{www.oasis-open.org} is consistent with NoPs.}\tabularnewline
\hline 
WG & Workgroups & WG are groups of technical experts whose goals span a business area or array of
organisational factors. WGs are always accompanied by a number of organisational sponsors and are expected to generate benefits as wide as their goals. For example, IFIP\footnote{\url{www.ifip.org}} working groups are WGs by definition.\tabularnewline
\hline 
PT & Project Teams & PTs are made by people with complementary skills who work together
to achieve a common purpose for which they are accountable. They are enforced by their organisation and follow specific strategies or organisational guidelines (e.g. time-to-market, effectiveness, low-cost). Their final goal is delivery of a product or service that responds to provided requirements \cite{ossslr}.\tabularnewline
\hline 
SC & Strategic Communities & SCs consist of meticulously selected people, experts in certain sectors
of interest to a corporation or or a set of organisational partners tied with formal non-disclosure agreements. These people attempt to proactively solve problems within strategic business areas of the organisational sponsor. Consultancy groups are often formed following the organisational theory around strategic communities.\tabularnewline
\hline 
FG & Formal Groups & FG members are screened and grouped formally by corporations. For example, a Learning Community can be structured as a FG if the group needs to undergo formal training and hands-on activities. Or such groups may be formed due to their special skills (e.g., the European Space Agency regularly puts together groups of ``Tiger Teams": skilled, experienced problem solvers to address mission-critical goals).  FGs have a single organisational goal, often called ``mission". In comparison to Formal Networks, they seldom rely on networking technologies. On the contrary, they
are local in nature.\tabularnewline
\hline 
PSC & Problem-solving Communities & PSCs are specific instances of Strategic Communities focused
on a solving particular problem. For example, Special-Interest Groups (SIGs) within ACM\footnote{\url{https://www.acm.org/special-interest-groups}} constitute strategic communities.\tabularnewline
\hline 
LC & Learning Communities & LCs provide a space for pure learning and explicit sharing of actionable
knowledge (i.e. skills). In a learning community, the leadership exercises membership approval and is tied to the learning objectives given to the member. Each developed or exchanged practice must become part
of the organisational culture \cite{Ruuska2003}. For example, StackOverflow is as much an example of informal community as is an example of Learning Community.\tabularnewline
\hline 
KC & Knowledge Communities &  KCs are groups of people with a shared passion to create, use, and share new knowledge for tangible business purposes (e.g.
increased sales, increased product offerings, client profiling).
The main difference with other types is that KCs are expected (by
the corporate sponsors) to produce actionable knowledge (that can be put to immediate use e.g. best-practices, standards, methodologies, approaches, problem-solving-patterns) in a
 business area \cite{ErikAndriessen2005}. For example, the Kaggle\footnote{\url{https://www.kaggle.com/}} community is a knowledge community around data science.\tabularnewline
\hline 
SN & Social Networks & SNs are emergent networks of social ties spontaneously arising between individuals who share a practice or common interest on a problem. SNs are a gateway for communicating communities \cite{Cross2005}.\tabularnewline
\hline 
\end{tabular}
\end{table*}

\subsection{Software Architectures}\label{arch}

Software architecture is an abstraction of a software system: its components, their properties, their relationships, as well as the choices that led to the system \cite{kazmanbook}. In this article we use architecture issues and  smells, reported by practitioners, as a way to understand and characterise the organisational structures under study. We designed our study to minimize bias as much as possible and thus we let our practitioners describe their projects' architectures, using reference models from the Microsoft Application Architecture Guide (MAAG), freely available on MSDN\footnote{\url{https://msdn.microsoft.com/en-us/library/ff650706.aspx}} and previously known to all practitioners involved in our study.  In the scope of this study, we used the architecture issues or ``smells" described in the MAAG, since: (a) practitioners used them to report which issues were present in their architectures; (b) we used these to evaluate organisational structures, characteristics, and relations found. The list of issues follows below:\footnote{Note that the terms used here are our own descriptive names for the issues reported in the MAAG.}

\begin{enumerate}
\item \textbf{Impossible Component Swap:} this issue reflects architecture components which are too tightly connected to the rest of the architecture which may lead to substitution problems.
\item \textbf{Untraceable Business Requirement:} this issue reflects overly fine-grained architectures where it becomes difficult to trace high-level business requirements across the entire architecture. This weighs on the ability to analyze the quality aspects behind that requirement and consequently the overall quality of the architecture.
\item \textbf{Sloppy Modularisation:} this issue reflects carelessly modularised architectures, e.g., architectures that randomly decrease cohesion for no good reason. This weighs heavily on the architecture's evolvability, its usefulness as a knowledge conveyor, as well as a means for division of work.
\item \textbf{Unscalable Architecting:} this issue reflects an architecture that can barely perform its functions in the face of rapidly increasing service demands; the consequent issues on architecture quality are obvious.
\item \textbf{Inflexible Architecture / God Classes:} this issue reflects architectures where a small number of components or modules contain significant functionality, creating bottlenecks for maintainability.
\item \textbf{Unmodifiable Core:} this issue reflects architectures that have gradually grown out of a core which is more and more becoming functionally untouchable and incomprehensible. This is often a consequence of Sloppy Modularisation.
\item \textbf{Spike-Centric Architecture:} this issue reflects architectures emerged directly from \emph{architecture spikes}, i.e., ``[...] architectural spike is a test implementation of a small part of the application's overall design or architecture". Relying heavily on architecture spikes may cause severe integration and evolution issues.
\item \textbf{Quality-Implicit Architecture:} this issue reflects architectures for which quality analysis of one (or more) software architecture properties (e.g., performance, reliability, safety) is so complex that it is never actually done.
\item \textbf{Architecture Monolith:} this issue reflects architectures whose design has grown into a fully connected mesh and therefore cannot be evolved other than by incurring heavy refactoring costs. Such monoliths may have originated as God Classes.
\item \textbf{Insensitive Information Spreading:} this issue reflects architectures in which one or more components carelessly disseminate the data that the software architecture is manipulating.  This can cause security vulnerabilities as well as privacy-policy violations.
\end{enumerate}

Finally, in this paper, we define the \emph{architectural debt} of an organisational structure pattern X as the rudimentary count of software architecture structural flaws and issues (selected from the list above) corresponding to that pattern: 

\begin{center}
\emph{Architectural debt:\\}
$\chi = \sum A_i(a_1,...a_n)$;
\end{center}

where $\chi$ represents the architectural debt of an organisational structure pattern while A$_i$ represents reported software architecture issues for projects $a_1,...a_n$ which are manifesting that pattern.  For the purpose of quantifying the relation between organisational structure patterns and software architectures, we are interested in measuring this quantity for all organisational structure patterns.

\subsection{Goals and Approach Summary}
We aim to determine if there exists a relationship between the characterisation of an organisational structure (established by simply counting its most evident characteristics) with respect to the characterisation of its underlying software architecture (established by simply counting its most evident issues). Returning to our original research questions: (1) \emph{is there a recurrent organisational structure pattern in agile software engineering teams?}; (2) \emph{if so, what does that pattern imply, in terms of software architecture quality?} We therefore derive a set of concrete research goals as follows:

\begin{enumerate}
\item[A] Understanding recurrent organisational characteristics: (A1.1) How frequently do organisational characteristics and types occur across a sample set of agile software organizations? (A1.2) How are the characteristics combined/clustered?
\item[B] Understanding architectural issues: how frequently do architectural issues occur?
\item[C] Understanding relations between A \& B: what is the correlation between phenomena in A and B?
\end{enumerate} 

Note that the adoption of a count for characteristics and architecture issues/smells follows a simple reasoning: it is the most basic approach to aggregate a measure of said characteristics in the scope of our study. It should be noted that this approach accounts for our exploratory intent wherefore our aim is to understand, in the most unadulterated way possible, the basic relation (if any) between organisational structures and software architectures.

\begin{table*}
\centering
\caption{Organisational Structure Types and Their Characteristics; every type (rows) can be identified by ascertaining the evident presence of the respective characteristics (columns).}\label{outline}
\tiny
\begin{tabular}{|>{\raggedright}p{1cm}|>{\raggedright}p{0.4cm}|>{\raggedright}p{0.4cm}|>{\raggedright}p{0.4cm}|>{\raggedright}p{0.4cm}|>{\raggedright}p{0.4cm}|>{\raggedright}p{0.4cm}|>{\raggedright}p{0.4cm}|>{\raggedright}p{0.4cm}|>{\raggedright}p{0.4cm}|>{\raggedright}p{0.4cm}|>{\raggedright}p{0.4cm}|>{\raggedright}p{0.4cm}|>{\raggedright}p{0.4cm}|}
\hline 
~~~~~~~~\textbf{Type}
\tiny
\emph{Char.} & \textbf{CoP} & \textbf{IN} & \textbf{FN} & \textbf{IC} & \textbf{NoP} & \textbf{WG} & \textbf{PT} & \textbf{SC} & \textbf{FG} & \textbf{PSC} & \textbf{LC} & \textbf{SN} & \textbf{KC}\tabularnewline
\hline 
\emph{Structure} & {\scriptsize{}high} & {\scriptsize{}high} & {\scriptsize{}high} & {\scriptsize{}high} & {\scriptsize{}high} & {\scriptsize{}high} & {\scriptsize{}high} & {\scriptsize{}high} & {\scriptsize{}high} & {\scriptsize{}high} & {\scriptsize{}high} & {\scriptsize{}high} & {\scriptsize{}high}\tabularnewline
\hline 
\emph{Situatedness} & {\scriptsize{}high} & {\scriptsize{}low} & {\scriptsize{}low} & {\scriptsize{}low} & {\scriptsize{}low} & {\scriptsize{}high} & {\scriptsize{}high} & {\scriptsize{}low} & {\scriptsize{}high} & {\scriptsize{}low} & {\scriptsize{}low} & {\scriptsize{}low} & {\scriptsize{}low}\tabularnewline
\hline 
\emph{Dispersion} & {\scriptsize{}low} & {\scriptsize{}high} & {\scriptsize{}high} & {\scriptsize{}high} & {\scriptsize{}high} & {\scriptsize{}low} & {\scriptsize{}low} & {\scriptsize{}high} & {\scriptsize{}low} & {\scriptsize{}high} & {\scriptsize{}low} & {\scriptsize{}low} & {\scriptsize{}high}\tabularnewline
\hline 
\emph{Problem-Focus} & {\scriptsize{}low} & {\scriptsize{}low} & {\scriptsize{}low} & {\scriptsize{}low} & {\scriptsize{}low} & {\scriptsize{}high} & {\scriptsize{}high} & {\scriptsize{}high} & {\scriptsize{}high} & {\scriptsize{}high} & {\scriptsize{}low} & {\scriptsize{}low} & {\scriptsize{}low}\tabularnewline
\hline 
\emph{Informality} & {\scriptsize{}high} & {\scriptsize{}high} & {\scriptsize{}low} & {\scriptsize{}low} & {\scriptsize{}high} & {\scriptsize{}high} & {\scriptsize{}high} & {\scriptsize{}low} & {\scriptsize{}high} & {\scriptsize{}high} & {\scriptsize{}high} & {\scriptsize{}low} & {\scriptsize{}high}\tabularnewline
\hline 
\emph{Formality} & {\scriptsize{}low} & {\scriptsize{}low} & {\scriptsize{}high} & {\scriptsize{}low} & {\scriptsize{}low} & {\scriptsize{}high} & {\scriptsize{}high} & {\scriptsize{}high} & {\scriptsize{}high} & {\scriptsize{}low} & {\scriptsize{}low} & {\scriptsize{}low} & {\scriptsize{}high}\tabularnewline
\hline 
\emph{Engagement} & {\scriptsize{}high} & {\scriptsize{}high} & {\scriptsize{}low} & {\scriptsize{}high} & {\scriptsize{}low} & {\scriptsize{}high} & {\scriptsize{}high} & {\scriptsize{}low} & {\scriptsize{}low} & {\scriptsize{}high} & {\scriptsize{}high} & {\scriptsize{}low} & {\scriptsize{}low}\tabularnewline
\hline 
\emph{Cohesion} & {\scriptsize{}low} & {\scriptsize{}low} & {\scriptsize{}high} & {\scriptsize{}low} & {\scriptsize{}low} & {\scriptsize{}high} & {\scriptsize{}high} & {\scriptsize{}low} & {\scriptsize{}low} & {\scriptsize{}low} & {\scriptsize{}low} & {\scriptsize{}low} & {\scriptsize{}low}\tabularnewline
\hline 
\emph{Duration} & {\scriptsize{}high} & {\scriptsize{}high} & {\scriptsize{}high} & {\scriptsize{}high} & {\scriptsize{}high} & {\scriptsize{}high} & {\scriptsize{}low} & {\scriptsize{}high} & {\scriptsize{}high} & {\scriptsize{}high} & {\scriptsize{}high} & {\scriptsize{}low} & {\scriptsize{}low}\tabularnewline
\hline 
\emph{ROI-Tracking} & {\scriptsize{}low} & {\scriptsize{}low} & {\scriptsize{}low} & {\scriptsize{}low} & {\scriptsize{}low} & {\scriptsize{}low} & {\scriptsize{}high} & {\scriptsize{}high} & {\scriptsize{}low} & {\scriptsize{}low} & {\scriptsize{}low} & {\scriptsize{}low} & {\scriptsize{}low}\tabularnewline
\hline 
\emph{Governance} & {\scriptsize{}low} & {\scriptsize{}low} & {\scriptsize{}high} & {\scriptsize{}low} & {\scriptsize{}low} & {\scriptsize{}high} & {\scriptsize{}high} & {\scriptsize{}high} & {\scriptsize{}high} & {\scriptsize{}low} & {\scriptsize{}low} & {\scriptsize{}low} & {\scriptsize{}low}\tabularnewline
\hline 
\emph{Culture-Tracking} & {\scriptsize{}low} & {\scriptsize{}low} & {\scriptsize{}low} & {\scriptsize{}low} & {\scriptsize{}low} & {\scriptsize{}high} & {\scriptsize{}low} & {\scriptsize{}low} & {\scriptsize{}low} & {\scriptsize{}high} & {\scriptsize{}high} & {\scriptsize{}low} & {\scriptsize{}low}\tabularnewline
\hline 
\emph{Visibility-Tracking} & {\scriptsize{}low} & {\scriptsize{}low} & {\scriptsize{}low} & {\scriptsize{}low} & {\scriptsize{}low} & {\scriptsize{}low} & {\scriptsize{}low} & {\scriptsize{}low} & {\scriptsize{}low} & {\scriptsize{}low} & {\scriptsize{}low} & {\scriptsize{}low} & {\scriptsize{}high}\tabularnewline
\hline 
\end{tabular}
\end{table*}

\section{Research Design}\label{rd}

To obtain evidence for this research we used a mixed-methods approach featuring 3 phases of data collection and quality-assessment as well as 2 subsequent analysis phases: (1) survey design and Delphi study \cite{gnatzy2011delphi}; (2) online survey; (3) confirmatory interviews; (4) data summary and observer reliability assessment; (5) content analysis. This section outlines our research design in detail, starting from a walkthrough of our theoretical framework, target population, and sampling strategy.

\subsection{Theoretical Framework}

\begin{figure}
\begin{center}
\includegraphics[scale=0.58]{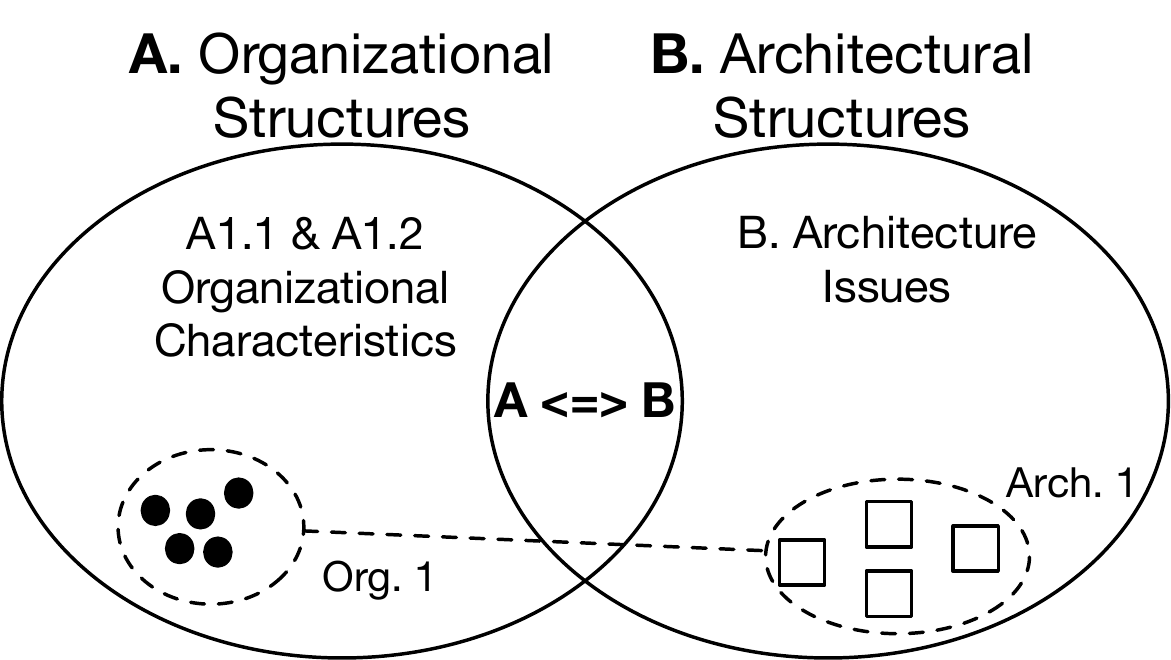}
\end{center}
\caption{A theoretical framework for our study - A. organisational structures (left-hand side) are graphs consistent with organisational characteristics; similarly, B. architectural structures (right-hand side) are graphs consistent with architecture issues; we are interested in characterising A (by uncovering A1.1 and A1.2), understanding the recurrence of B and finding/quantifying the relations between A and B.}\label{theor}
\end{figure}

Figure \ref{theor} outlines the theoretical framework behind this study. On one hand, organisational structures (marked A, on the left-hand side) are graphs consistent with organisational characteristics. Similarly, architectural structures (marked B, on the right-hand side) are graphs consistent with architecture issues. We are interested in characterising A (by uncovering A1.1 and A1.2), understanding the recurrence of B and finding/quantifying the relations between A and B.

To address our theoretical framework we employ: (A1.1) organisational and frequency analysis combined with descriptive statistics; (A1.2) graphs and clusters analysis; (B) architecture and frequency analysis combined with descriptive statistics; (A$<$=$>$) descriptive statistics and correlation analysis.

\subsection{Target Population and Sampling Strategy}

Our study covered more than 4 years, involving 90 people in 9 different organizations and arranged in 30 different teams (with an average of 3 teams per organization). To attain this considerable sample size, we adopted the following process. Invitation letters were sent over this time-window to 136 practitioners  with 4+ years of experience with software and its engineering. Following a strategic sampling strategy \cite{0029933}, the initial set of practitioners were from our own networks: developers and architects who had previously worked with one or more of the authors of this study.  Practitioners were initially contacted with an email asking of their interest in the study and, if interested, were asked for contact details for 2 additional colleagues (i.e., a \emph{snowballing rule} \cite{snowball}) who might be available for the study as well. The process of establishing agility was built-in within our invitation letter - in particular, we employed the practices defined in previous work \cite{Meyer14} to map the practitioners onto agile practices.

Out of our initially-sampled 136 practitioners, 42 responded and 12 were filtered out as not feasible. Our control factors were aimed at covering a sufficiently generalisable sample, diversifying:
\begin{itemize}
\item codebase size: split evenly among medium-sized (200-500 KLOC) projects, 33\% large (500-850 KLOC) projects, and 33\% very large ($>$ 850 KLOC) projects; 
\item main programming language - Java, C\#, C, Python. In addition YAML and other scripting languages are included for projects in our sample;
\item team size - our size distribution is evenly split among three ranges: small to medium ($<$7 members), large (8$>$15 members) and very-large ($>$15 members);
\item team age - our project team age distribution is evenly split among three
ranges: young ($<$24 months), established (24$>$32 months), and
mature ($>$32 months);
\item architecture type - all our projects were sampled choosing service-oriented or service-based applications determined using concepts and definitions from literature \cite{soa} --- this design choice is connected to the reported proneness of certain architecture styles for certain architecture issues \cite{kazmanbook,Kazman15debt};
\end{itemize}

As a result of the above filtering a final sample size of 30 starting practitioners was obtained. Following our snowballing rule, the population for our study involved therefore a total of 90 practitioners from 9 different organizations arranged in a total of 30 teams.  The market segments for the 9 organizations represented are: aerospace, heavy automotive industry, mobile-phone manufacturing, information systems consulting (two organizations), healthcare informatics, banking information systems, food production, and electronics.

\subsection{Phases 1 and 2 - Survey Design and Execution}\label{desex}

In this study, we aimed at a defensible and replicable qualitative-quantitative research design, striving to minimise our own interpretation of raw data directly elicited from the involved practitioners. In so doing, we hard-coded a Delphi-inspired approach \cite{gnatzy2011delphi} to refine a survey questionnaire which would directly and reproducibly yield the data we sought. In particular, a Delphi study relies on a panel of experts to answer questionnaires in two or more rounds (two rounds, in our case). In our case, after a first round of survey (Phases 1 and 2), a facilitator provided an anonymised summary of the previous two experts' forecasts.  The third practitioner (Phase 3, see Sec. \ref{conf}) was then encouraged to confirm, deny, or revise answers in light of the replies of other members of their panel. As previously stated, a datapoint for Table 5 and 6 was obtained only if {\em all} the three practitioners agreed that a certain architecture issue, or organisational characteristic was present. 

\begin{figure}
\begin{center}
\includegraphics[scale=0.58]{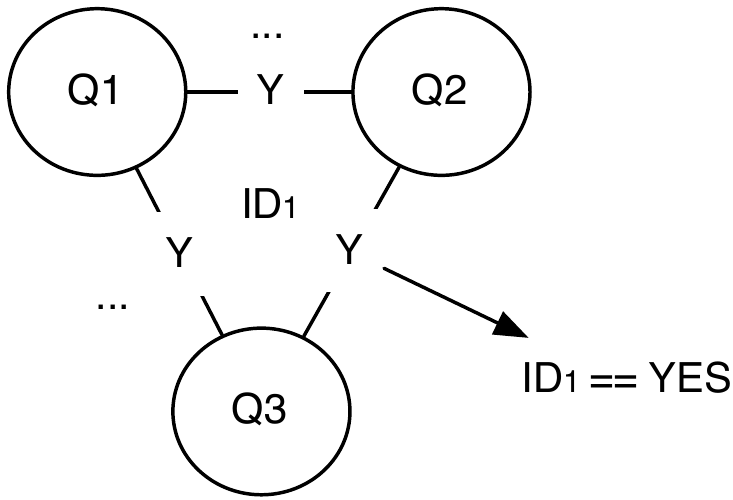}
\end{center}
\caption{A Voting System For Question Triangulation (Q1,2,3) and Identifier Elicitation (ID$_1$); three questions reflect the same identifier/characteristic, a voting system is used to understand if the characteristic is prominent or not. }\label{voting}
\end{figure}

To design the survey for this study, we operated as follows.
First, to elicit organisational structure patterns in the target population, we prepared a list of questions that reflected 12 organisational structure pattern \emph{identifiers}, that is, the key organisational or socio-technical characteristics (e.g., member selectivity, informal communication) that define an organisational structure type, based on definitions from the state of the art \cite{ossslr}. 
Each structure identifier was associated with a set of 3 closed, ternary (Yes-No-Maybe/Unknown), and equivalent (i.e., phrased to reflect the same attribute) questions. This research design feature was inspired by Delphi studies and enabled us to: (a) adopt a voting system to infer the presence/absence of structure identifiers (see Fig. \ref{voting} for an overview of the technique), meaning that a positive elicitation (YES) of a structural identifier is determined if at least 2 questions are responded positively, and negative in all other cases; (b) automatically triangulate responses at the source \cite{Svensson2012}. The questionnaire thus distilled is available online.\footnote{\url{http://tinyurl.com/ly9hoof}}

Second, we prepared a study summary and invitation letter containing: (a) links to our organisational elicitation survey; (b) links to our organisational issues and community smells identification survey\footnote{While further discussion of community smells elicitation is out of the scope of this article, more details can be found in \cite{archcomm}.}; (c) links to the Microsoft Application Architecture Guide (MAAG), with a request to report back via email a 1-line description of their architecture using patterns from MAAG as well as any and all architecture issues from the MAAG that they observed in their own architecture. This immediate email response served the purpose of avoiding any bias introduced by the authors regarding architecture issues.

We aimed to survey 2 out of 3 practitioners belonging to the same project. The 3rd practitioner was subsequently interviewed directly for confirmatory purposes. Furthermore, because the scope of our study aims at offering the most certain and confirmed presence of architecture smells and organisational characteristics, we only acknowledged the existence of an organisational characteristic, an organisational type, or an architecture smell, if and only if both the surveyed practitioners *and* the third interviewed practitioner all agreed and indicated the presence of the characteristic, or architecture smell (see Table 4 and 5 towards the end of the manuscript). This interpretation of our methodology was employed to avoid any wild assumptions behind the data or the involved quantities - rather, this interpretation of the research method aims at giving our data a confirmatory connotation. Conversely, we are planning subsequent studies in which we aim to further understand the quantities behind our research subjects---these are delayed to future work.

\subsection{Phase 3 - Confirmatory Interviews}\label{conf}

To instrument phase 3 of our study, we reached back to the previously uninvolved practitioner for every project in our sample. These practitioners were interviewed with a variation of our complete survey questionnaire (see Sec. \ref{desex}) intended to confirm all the observations of their colleagues about their team as well as clarify any clearly conflicting answers. For this phase, we adopted an interview guide approach.\footnote{interview guide available here: \url{http://tinyurl.com/kl97cgq}}  At the conclusion of each interview we asked interviewees to fill in the organisational pattern detection algorithm and questionnaire defined previously, to further confirm the validity of our findings. Thus structured, phase 3 led to a total of 31 interviews amounting to over 50+ hours of recorded material.  While we cannot disclose the raw data from survey Phases 1 to 3, we have prepared a spreadsheet containing aggregated data from responses and confirmatory interviews. To encourage verifiability, this dataset is freely available.\footnote{\url{https://tinyurl.com/y7q96uy8}}

\subsection{Data Synthesis and Reliability Assessment}\label{synthesis}

All available data was loaded into tables formatted to be analysed with the R data analytics toolkit in parallel by two of the authors of the paper; the purpose of analysis was to produce the statistics, plots and representations currently showcased in Sec. \ref{res}. At this point, we assessed observer reliability evaluating the agreement across analyses over the data tables we produced. Subsequently, we computed the Krippendorff's $\alpha$ coefficient for observation agreement \cite{krippendorff04} - the $\alpha$ score essentially measures a confidence interval score stemming from the agreement of values across two distinctly-reported observations about the same event or phenomenon. In our case the value was applied to measure the agreement between configuration details, analyses, and statistical results of our analysis. The value was calculated initially to be 0.83, hence $\alpha>.800$, which is a standard reference value for highly-confident observations.  Subsequently, the value was used to drive the agreement between the two analyses up towards total alignment.

\subsection{Data Analysis}

To analyse the results in this paper from a statistical perspective (i.e., checking for normalization and evaluating the risks of multicollinearity \cite{Ratner2009}), we prepared descriptive statistical plots (see Figures from \ref{orgchar} to \ref{orgpat}) for our data. Subsequently, having established the normal distribution of the data features to be correlated further, we computed Pearson correlation calculations across the following dimensions:
\begin{itemize}
\item occurrence of reported architecture smells and the occurrence of prominent organisational characteristics (i.e., our own organisational measure);
\item team/community age and reported numbers of architecture issues;
\item team/community size and reported numbers of architecture issues;
\end{itemize}

The use of Pearson product-moment correlation is well-formed since our study design ``flattens" the quantities involved (i.e., the magnitude of organisational characteristics and of individual architecture smells) to linear sums which are more appropriately correlated by means of product-moment analysis \cite{FouladiS08}.
Furthermore, we analysed the reported structural identifiers by means of the decision-tree we defined and evaluated in \cite{specissue}. The decision-tree encodes the set of relations (e.g., implication or mutual-exclusion) across primary identifiers from Figure \ref{outline}. This set of relations forms, by definition, a partial-order function, i.e., a function that associates an ordering or sequencing to the elements of a set. The decision-tree (see Fig. \ref{tree}) is a representation of this partial-order function and is to be visited top-to-bottom (most generic to most specific type) and right-to-left (most collocated to most dispersed type).\footnote{All relations and decision-tree functional demonstration by construction can be found online at \url{http://tinyurl.com/mzojyp2}} With respect to the definitions provided previously, the tree is a mechanism to identify what types co-exist in an observed organisational structure by excluding all other possible combinations of types at the same time, by excluding the characteristics which are either opposite or inconsistent.\footnote{Further elaboration on the mechanics behind the decision-tree are available in the full technical report: \url{https://tinyurl.com/y6voory6}}

The output of decision-tree analysis is the set of organisational structure types (see Table \ref{overview}) emerging for each project. Two decision-tree analyses were initiated in parallel by two of the authors of this study. To strengthen the reliability of our observations and thus the validity of our results, we again conducted observer reliability assessment just like we did for survey responses reliability (see Sec. \ref{synthesis}) and computed once more the Krippendorff $\alpha$ coefficient for observation agreement between the parallel analyses. We reported an agreement value calculated to be 0.81, hence $\alpha>>.800$. Also, as a result of this observer reliability assessment step, one decision-tree visit was decided to be inconclusive (i.e., not leading to the identification of organisational structure characteristics and types, see project 17 on Table \ref{results} in the Appendix in the Appendix) and an additional set of 7 entries were marked as uncertain (see Table \ref{results} in the Appendix).

Finally, we used pattern-matching and recurrence analysis to identify most recurrent organisational structures, i.e., most frequent sets of types resulting from decision-tree visits (see Sec. \ref{patternsfinal}). For this analysis we exploited an instance of the K-repeating sub-string matching algorithm (KRS) \cite{MatsubaraH14} running on types elicited via decision-tree visits. KRS' main objective is to find the length of the longest substring T of a given string (the string of detected organisational structure types, in our case) such that every character in T appears no less than K times. Normally KRS is exploited to identify textual patterns for anonymisation purposes; in our context, it fits perfectly to the detection of recurring, non-overlapping (i.e., with at most one element in common) organisational structure patterns.  We used the organisational type IDs (see Table \ref{overview}) as seeds for matching. To run the algorithm, we used the Wolfram mathematical computing engine.

\section{Results}\label{res}

This section outlines our results, beginning with the descriptive statistics for our dataset. 

\subsection{Population Description}

\begin{figure*}
\centering
\includegraphics[scale=0.451]{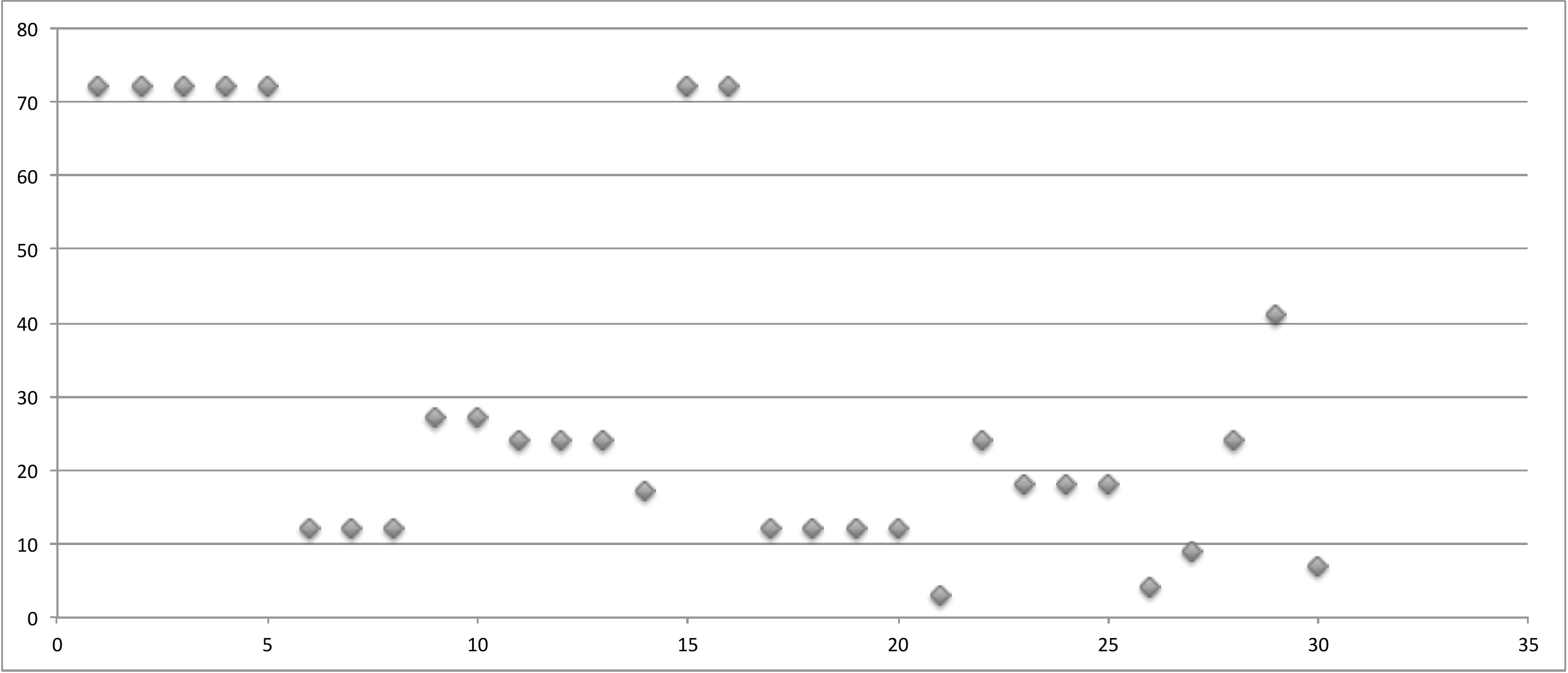}
\caption{Team Age Across the Dataset---the scatter-plot places the ID of each datapoint (from 1 to 30) in correspondence to its ``age", expressed in man-months since the creation of the organisational unit.}\label{teamage}
\end{figure*}

\begin{figure}
\centering
\includegraphics[scale=0.86]{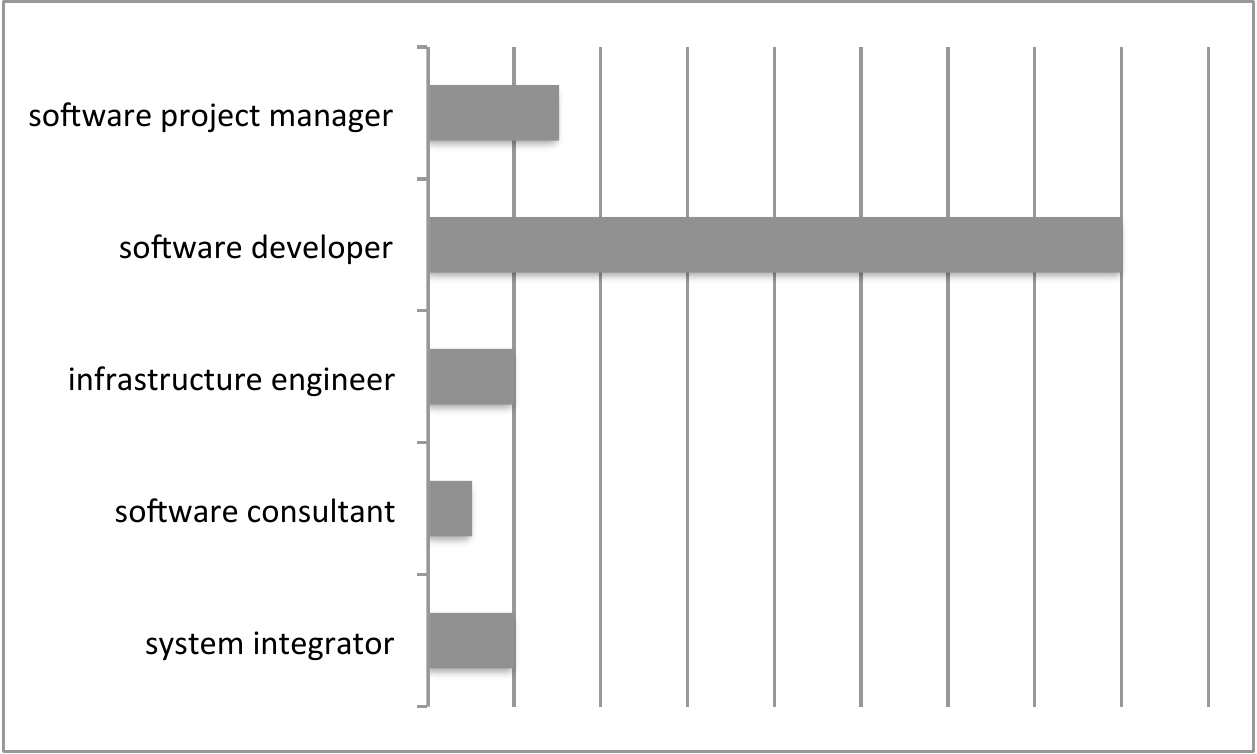}
\caption{Participant Roles Across the Dataset---the X-axis indicates frequency; the Y-axis indicates the role.}\label{role}
\end{figure}

As a result of our sampling strategy, our study was varied enough to contain a sufficient diversity of practitioner ages (see Fig. \ref{teamage}) and roles (see Fig. \ref{role}), with an average age of 45. An overview of every organisational structure under study is presented in Tab. \ref{tabbaoverview}. 

\begin{table*}[]
    \centering
\scriptsize
\begin{tabular}{|l|l|l|l|l|l|}
\hline 
\textbf{CI} & \textbf{Community Pattern} & \textbf{Model} & \textbf{Size} & \textbf{Age} & \textbf{\#issues}\tabularnewline
\hline 
7 & CoP,FN,IC,WG,SC,PSC,KC,LC & Scrum & 10 & 72 & 8\tabularnewline
\hline 
8 & CoP,FN,IC,WG,SC,PSC,SN & Scrum & 15 & 72 & 4\tabularnewline
\hline 
5 & FN,IC,PT,SC,PSC,SN & Scrum & 15 & 72 & 2\tabularnewline
\hline 
7 & IN,IC,WG,PSC,SN & Scrum & 15 & 72 & 6\tabularnewline
\hline 
4 & FN,NoP,PT,SC,PSC,SN & Agile-Waterfall & 15 & 72 & 7\tabularnewline
\hline 
8 & CoP,IN,IC,WG,PSC,SN,LC & Other Agile & 7 & 12 & 3\tabularnewline
\hline 
7 & CoP,IN,IC,WG,PSC,SN,LC,SC & Other Agile & 18 & 12 & 5\tabularnewline
\hline 
8 & CoP,IN,IC,WG,PSC,SN,FN & Other Agile & 18 & 12 & 5\tabularnewline
\hline 
6 & IN,IC,WG,PSC,SN,SC & Other Agile & 26 & 27 & 7\tabularnewline
\hline 
5 & CoP,IC,WG,PSC,SN,SC & Other Agile & 26 & 27 & 7\tabularnewline
\hline 
6 & CoP,IC,WG,PSC,SN & Other Agile & 5 & 24 & 4\tabularnewline
\hline 
7 & CoP,IN,IC,WG,PSC,SN,LC & Other Agile & 5 & 24 & 3\tabularnewline
\hline 
8 & IN,IC,FN,WG,PSC,SC,KC & Other Agile & 7 & 24 & 5\tabularnewline
\hline 
7 & IC,WG,PSC,SC,KC,FN & Scrum & 20 & 17 & 3\tabularnewline
\hline 
7 & IN,IC,FN,PSC,SN,LC & Scrum & 8 & 72 & 3\tabularnewline
\hline 
3 & - & Other Agile & 20 & 72 & 3\tabularnewline
\hline 
8 & CoP,IN,IC,WG,SN,FG & Scrum & 20 & 12 & 3\tabularnewline
\hline 
8 & IC,PT,PSC,SN,LC & Scrum & 16 & 12 & 4\tabularnewline
\hline 
5 & CoP,IC,WG,PSC,SN,SC & Scrum & 15 & 12 & 2\tabularnewline
\hline 
9 & CoP,IN,IC,WG,PSC,SN & Scrum & 27 & 12 & 2\tabularnewline
\hline 
9 & CoP,IN,IC,WG,PSC,SN,LC & Scrum & 9 & 3 & 2\tabularnewline
\hline 
9 & FN,IC,WG,PSC,LC,SN & Agile-Waterfall & 11 & 24 & 2\tabularnewline
\hline 
7 & CoP,IN,NoP,WG,SC,PSC,KC,LC & Other Agile & 13 & 18 & 2\tabularnewline
\hline 
8 & CoP,IN,NoP,WG,SC,PSC,KC,LC,FN & Other Agile & 13 & 18 & 3\tabularnewline
\hline 
9 & CoP,IN,IC,WG,SN,FG & Scrum & 13 & 18 & 3\tabularnewline
\hline 
10 & CoP,IN,IC,WG,SN,FG & Scrum & 4 & 4 & 3\tabularnewline
\hline 
6 & FN,IC,PT,SC,FG,KC & Other Agile & 13 & 9 & 3\tabularnewline
\hline 
7 & IC,PT,SC,FG,KC & Scrum & 12 & 24 & 4\tabularnewline
\hline 
10 & IN,IC,WG,PSC,KC,LC & Scrum & 12 & 41 & 4\tabularnewline
\hline 
8 & CoP,IN,NoP,PT,PSC,SN & Agile-Waterfall & 24 & 7 & 11\tabularnewline
\hline 
\end{tabular}
    \caption{Dataset overview; every datapoint reflects an organisational structure in our dataset starting with a characterisation index in column 1 (CI).}
    \label{tabbaoverview}
\end{table*}

The table outlines all 30 of our datapoints, outlining their organisational characterisation index, community pattern found, agile process model, community size \& age, as well as reported architecture issues. Concerning the process model dimension, as previously stated, we reported any differences by the claimed agile methods model either as ``other agile"---meaning any agile model other than Scrum--- or ``Scrum-Waterfall"---meaning when an agile methods approach was blended within an organization-wide classical waterfall model (e.g., in the scope of large-scale systems engineering companies).

\begin{figure}
\centering
\includegraphics[scale=0.86]{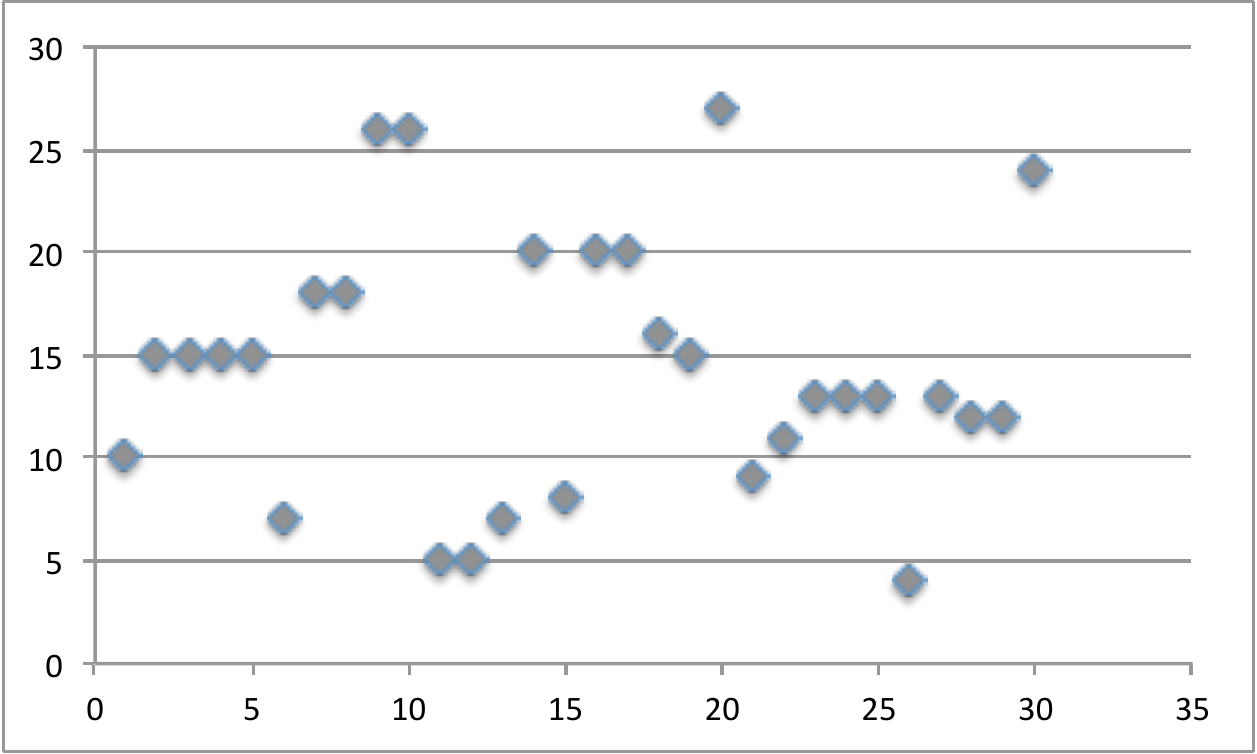}
\caption{Team Size Across the Dataset---the scatter-plot places the ID of each datapoint (from 1 to 30) in correspondence to its ``size", expressed in members of the entire organisational structure, which comprises multiple sub-teams.}\label{size}
\end{figure}

We were also able to achieve a reasonable diversity of team sizes (see Fig. \ref{size}, where the X-axis indicates our projects while the Y-axis indicates number of members), with an average size of 14.4 people and a standard deviation $\sigma = 6.21$. 

And we were able to evaluate a sufficiently diverse sample also in terms of \emph{team age}, i.e., the amount of time the team members have been working together on the same product, with an average age of 29.9 months and standard deviation $\sigma = 24.43$ (see Fig. \ref{teamage}, where the X-axis indicates our projects while the Y-axis indicates team age).  We were, however, unable to reach many female software professionals, registering just one female participant in this study.

\subsection{Organisational Characteristics}

\begin{figure*}
\centering
\includegraphics[scale=0.51]{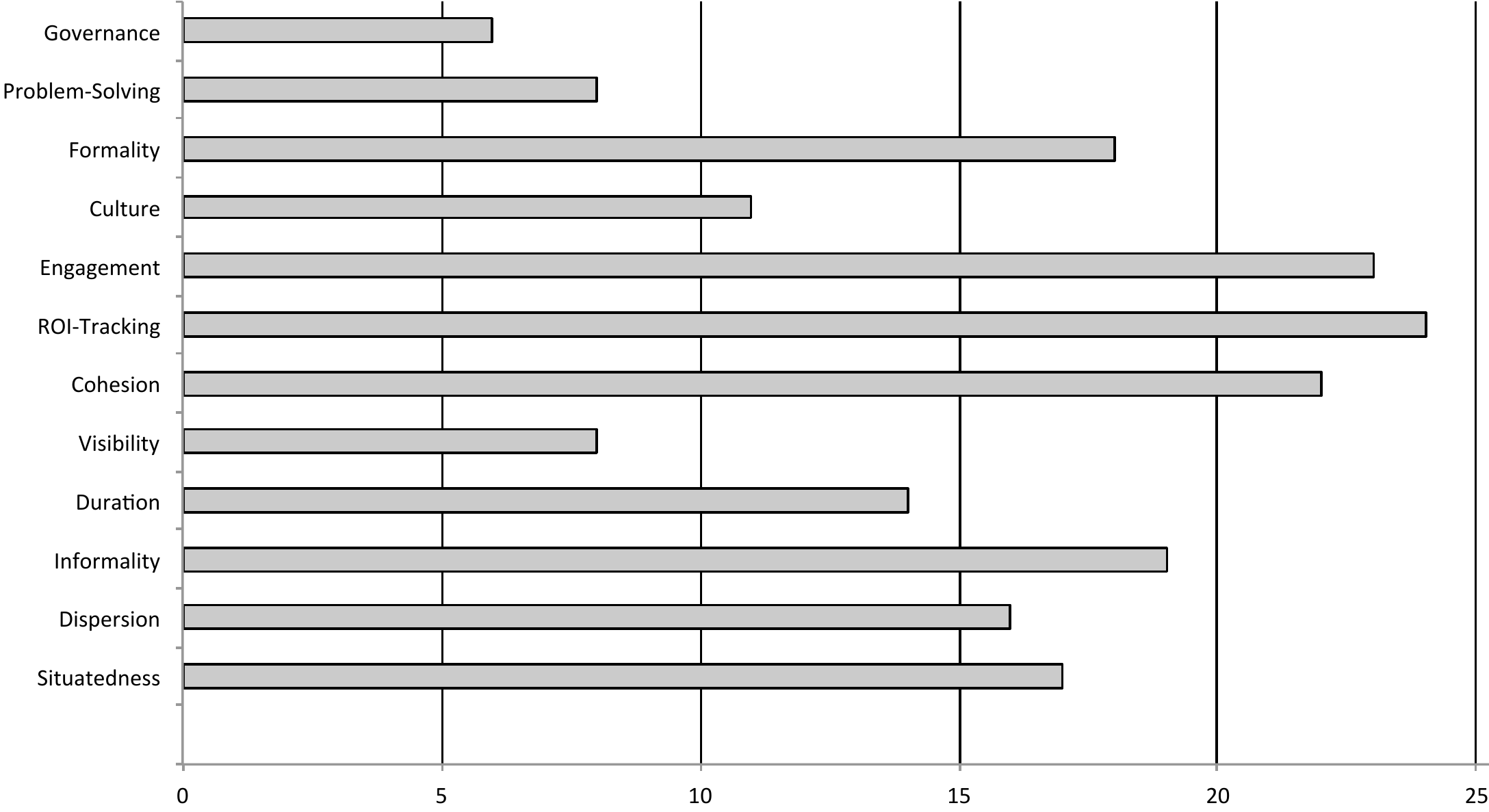}
\caption{Most Frequent Organisational Characteristics---X-axis reports frequency, while Y-axis reports the organisational characteristic.}\label{orgchar}
\end{figure*}

Figure \ref{orgchar} plots the most frequent organisational characteristics reported in our sample. The \emph{ROI-Tracking} characteristic is the most frequent. This characteristic denotes the importance of Return-On-Investment (ROI) as a key-performance indicator for project teams, and it was reported 20+ times.  This is not surprising, given that ROI has been a key motivator for software engineering efforts since the inception of software \cite{vanVliet03}.

The next most frequently observed characteristic is \emph{engagement}, which denotes an organisational precedence for personal engagement and self-organisation in work activities, rather than prescribed work. The \emph{cohesion} characteristic is next most frequent, and denotes tight collaboration and communication in a team. Subsequently we find the \emph{formality}, and \emph{informality} characteristics, which denote the presence or absence of formal protocols for knowledge/people interactions, retrieval, sharing, and protection across the organisation. Further on, we have \emph{situatedness}, denoting joint, collocated practice \cite{situatedness} and \emph{dispersion}, a practice where activities are geographically distributed over time and space. 

The set of characteristics shown on the Y axis of Figure \ref{orgchar} reflect the management practices reported by the teams in our study, such as the definition of a prescribed \emph{duration} for projects, or the presence of organisational \emph{culture}-tracking systems which were reported in 14 and 11 projects respectively.  Also, the presence of social or organisational \emph{visibility}-tracking tools was reported in 7 projects along with evidence indicating the explicit focus on a single technical problem or \emph{problem-solving} activity for their team. 

With respect to our correlation analysis, Table \ref{chars} reports the correlation between every individual organisational characteristic and architecture debt, as defined above. The analysis of correlations reveals that two characteristics (reported in bold), namely, \emph{cohesion} and \emph{culture-}tracking mechanisms are moderately correlated with the occurrence of architecture issues.

\begin{table}
\caption{Evaluation of individual organisational characteristics with respect to reported architecture issues - a moderate and meaningful correlation is reported for Cohesion and Culture-Tracking.}\label{chars}
\small
\centering
\begin{tabular}{|c|c|c|c|}
\hline 
 \textbf{Char.} & \textbf{Correl.} & \textbf{P-Value} & \textbf{Rel.} \tabularnewline
\hline 
Situatedness & -0,08504317 & $<<$0.01 & \tabularnewline
\hline 
Dispersion & 0,18149293 & 0,285140405 & \tabularnewline
\hline 
Informality & 0,097887993 & 0,380552469 & \tabularnewline
\hline 
Duration & -0,013971332 & $<<$0.01 & \tabularnewline
\hline 
Visibility-Tracking & -0,088729877 & $<<$0.01 & \tabularnewline
\hline 
\textbf{Cohesion} & \textbf{-0,20765179} & \textbf{$<<$0.01} & \textbf{X}\tabularnewline
\hline 
ROI-Tracking & 0,14349371 & 0,327388865 & \tabularnewline
\hline 
Engagement & -0,072597172 & $<<$0.01 & \tabularnewline
\hline 
\textbf{Culture-Tracking} & \textbf{-0,301994871} & \textbf{$<<$0.05} & \textbf{X}\tabularnewline
\hline 
Formality & -0,044364939 & $<<$0.01 & \tabularnewline
\hline 
Problem-Focus & 0,022090616 & 0,472722548 & \tabularnewline
\hline 
Governance & -0,177543065 & $<<$0.01 & \tabularnewline
\hline 
\end{tabular}
\end{table}

\subsection{Organisational Patterns}

Table \ref{results} in the Appendix outlines the set of types reported across our entire dataset (columns 2 to 9) stemming from characteristics shown in Figure \ref{orgchar}. From our analysis we have observed every organisational structure type at least once. In addition each pattern exhibits a minimum of 5 and a maximum of 9 types (for example, see projects 15 and 25).

\begin{figure*}
\centering
\includegraphics[scale=0.46]{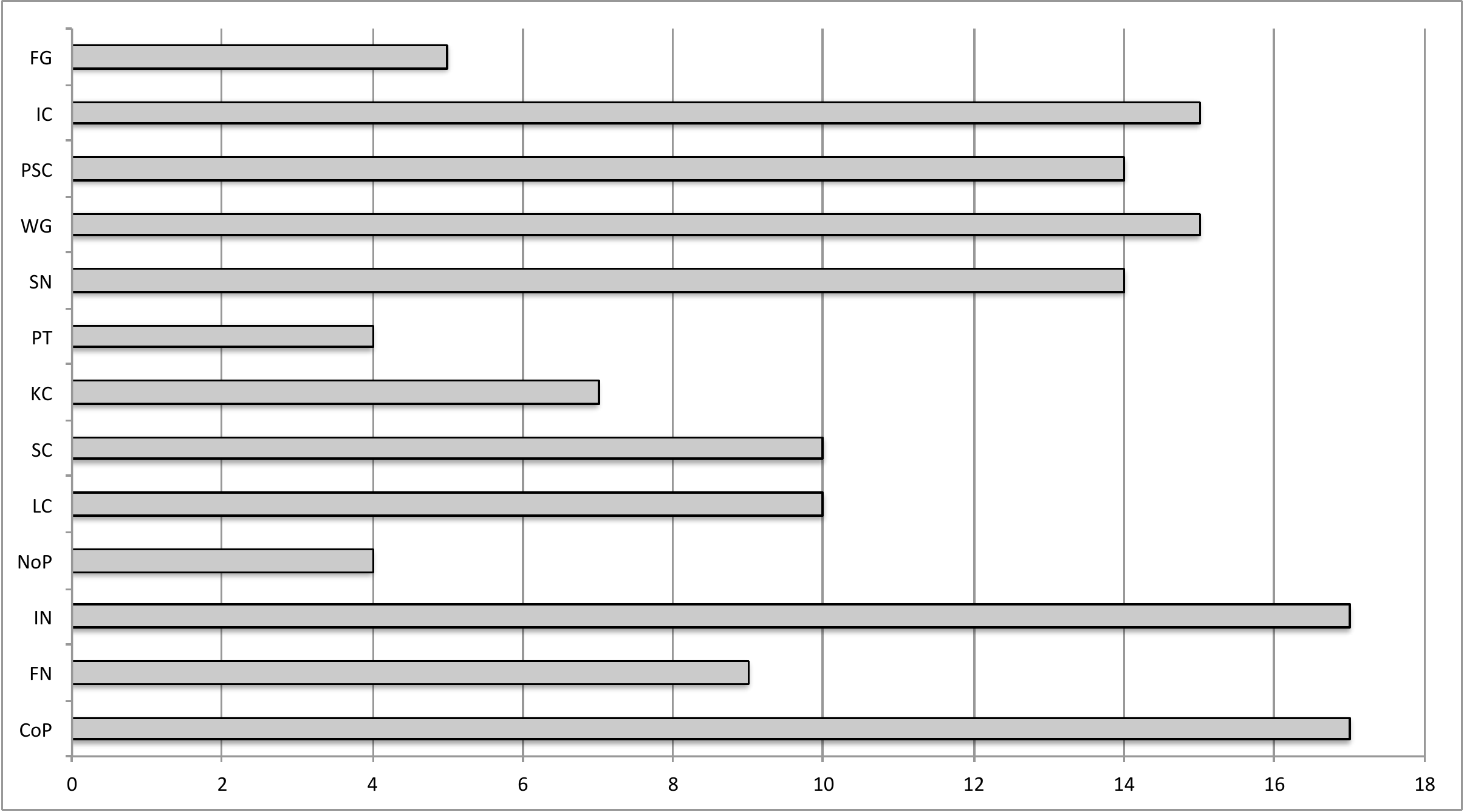}
\caption{Most Frequent Organisational Types Emerging from Characteristics in Fig. \ref{orgchar}---X-axis reports frequency while Y-axis reports the type.}\label{treetyping}
\end{figure*}

Figure \ref{treetyping} plots the frequencies of organisational structure types resulting from our decision-tree analysis.
The most predominant types are informal ones (CoP, IN, IC), those exhibiting a tightly knit/cohesive organisational structure (WG, PSC), or loosely-structured communities (SN,LC,KC).  Conversely, formal (FN,FG) or highly-structured and dispersed (NoP,SC) organisational structures are seen less frequently. These results are not surprising considering the agile methods adopted by the teams in our sample.  Agile methods promote informality, tight and cohesive collaborations, or loosely-structured communities (e.g. \cite{Polk11}).

\subsection{Architecture Issues}

\begin{figure*}
\centering
\includegraphics[scale=0.48]{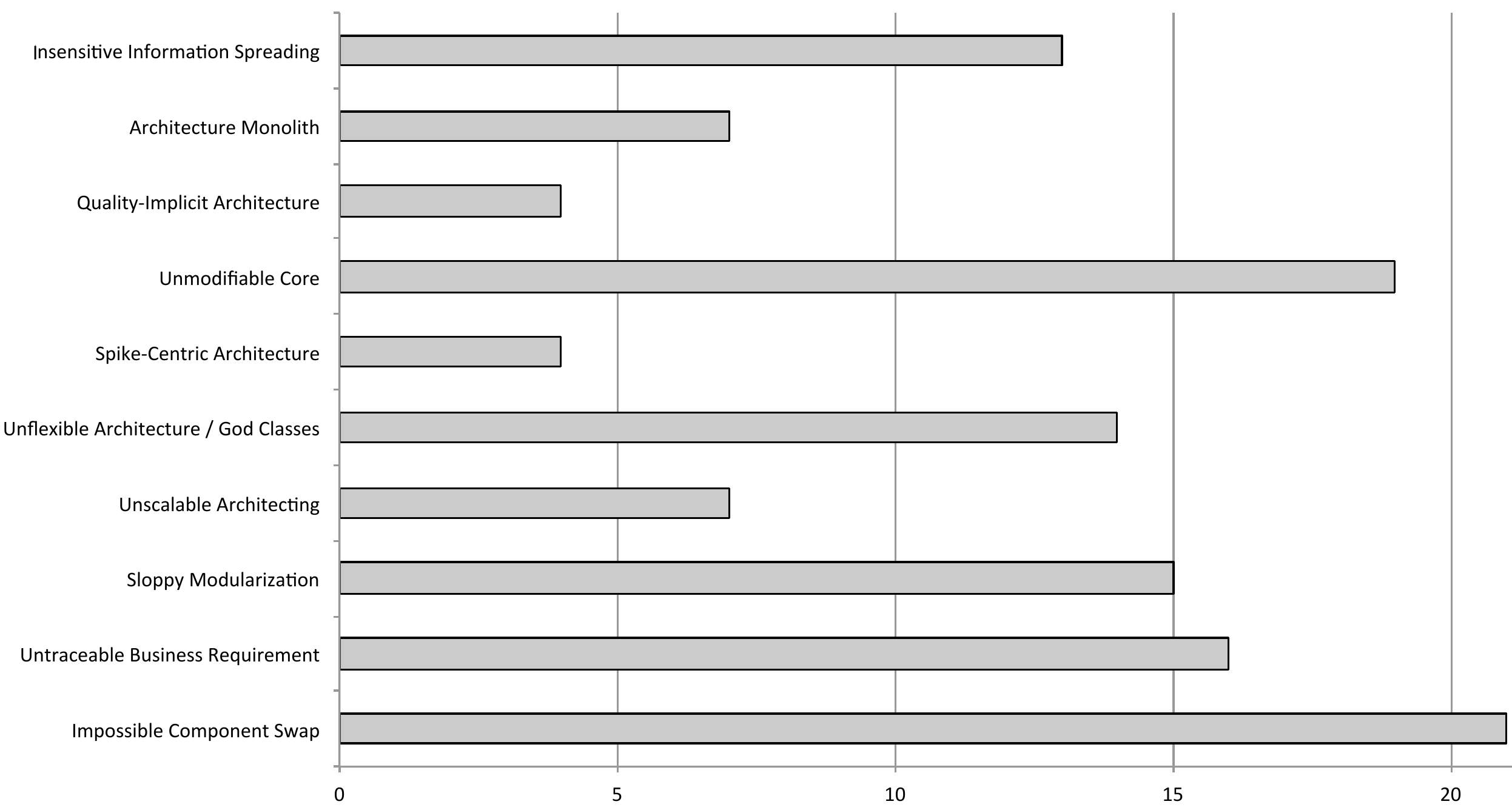}
\caption{Most Frequent Software Architecture Issues from our Dataset.}\label{archissues}
\end{figure*}

Figure \ref{archissues} plots the most frequent software architecture issues reported in our dataset, while Table \ref{issuesoverview} in the Appendix provides a mapping between projects and specific architecture smells. This figure highlights the predominance of unmodifiable or unsubstitutable architecture elements, followed by misguided business requirements (e.g., untraceable business requirements or insensitive information spreading). Subsequently, several modularisation issues are observed, for example, sloppy modularisation, architecture monoliths, or god classes. Issues connected to implicit or eroding architecture quality such as quality-implicit, spike-centric architecture, and unscalable architecture smells are seen least frequently.

\section{Discussion}\label{disc}

This section discusses our results and major contributions. We begin by outlining the organisational structure patterns we observed in our sample,  focusing on the most frequent.

\subsection{Qualitative Analysis: Organisational Structure Patterns from Agile Teams}\label{patternsfinal}

\begin{figure*}
\centering
\includegraphics[scale=0.42]{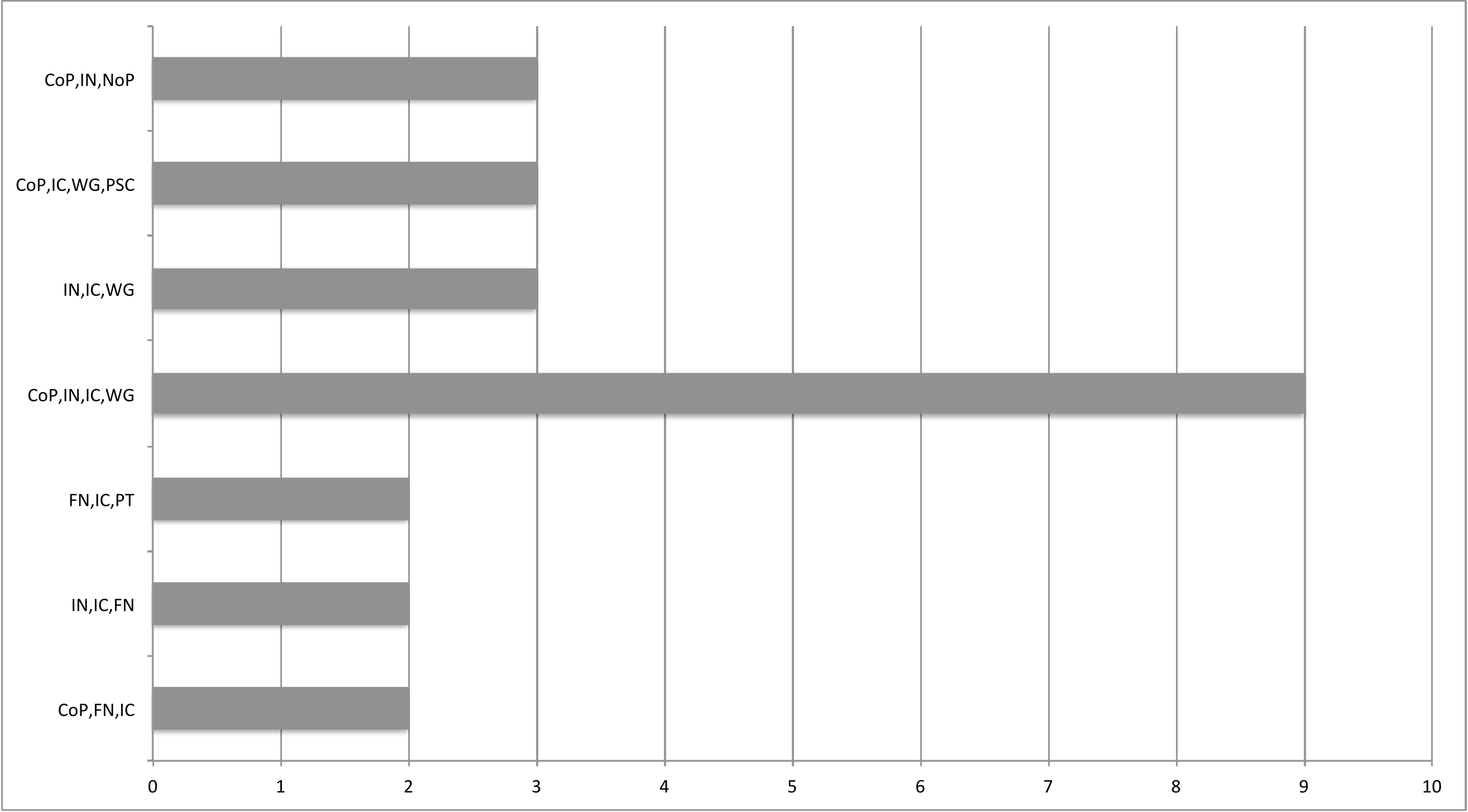}
\caption{Organisational Structure Patterns in Software Projects---the Y-Axis indicates the patterns while the X-axis contains \# recurrence; Patterns are ordered by increasing formality based on the presence of more formal organisational structures.}\label{orgpat}
\end{figure*}

Figure \ref{orgpat} outlines the recurrent patterns of organisational structure types we identified. 

\begin{framed}
\textbf{Finding 1.} According to our data, there are 7 non-overlapping patterns, consisting of at least 3 community types, that recur across our dataset.  Notably, a single pattern covers more than 37\% of our dataset.
\end{framed} 

In the following, we name and describe each of these structure patterns, interpreting them by means of models and characteristics from the literature \cite{ossslr}. Figures \ref{p1} to \ref{q5} illustrate the patterns found, with their salient characteristics annotated on nodes (i.e., members) and edges (i.e., interactions).

\subsubsection{P1: Formally-Networked Informal CoPs (CoP,FN,IC)} 
\begin{figure*}
\centering
\includegraphics[scale=0.5]{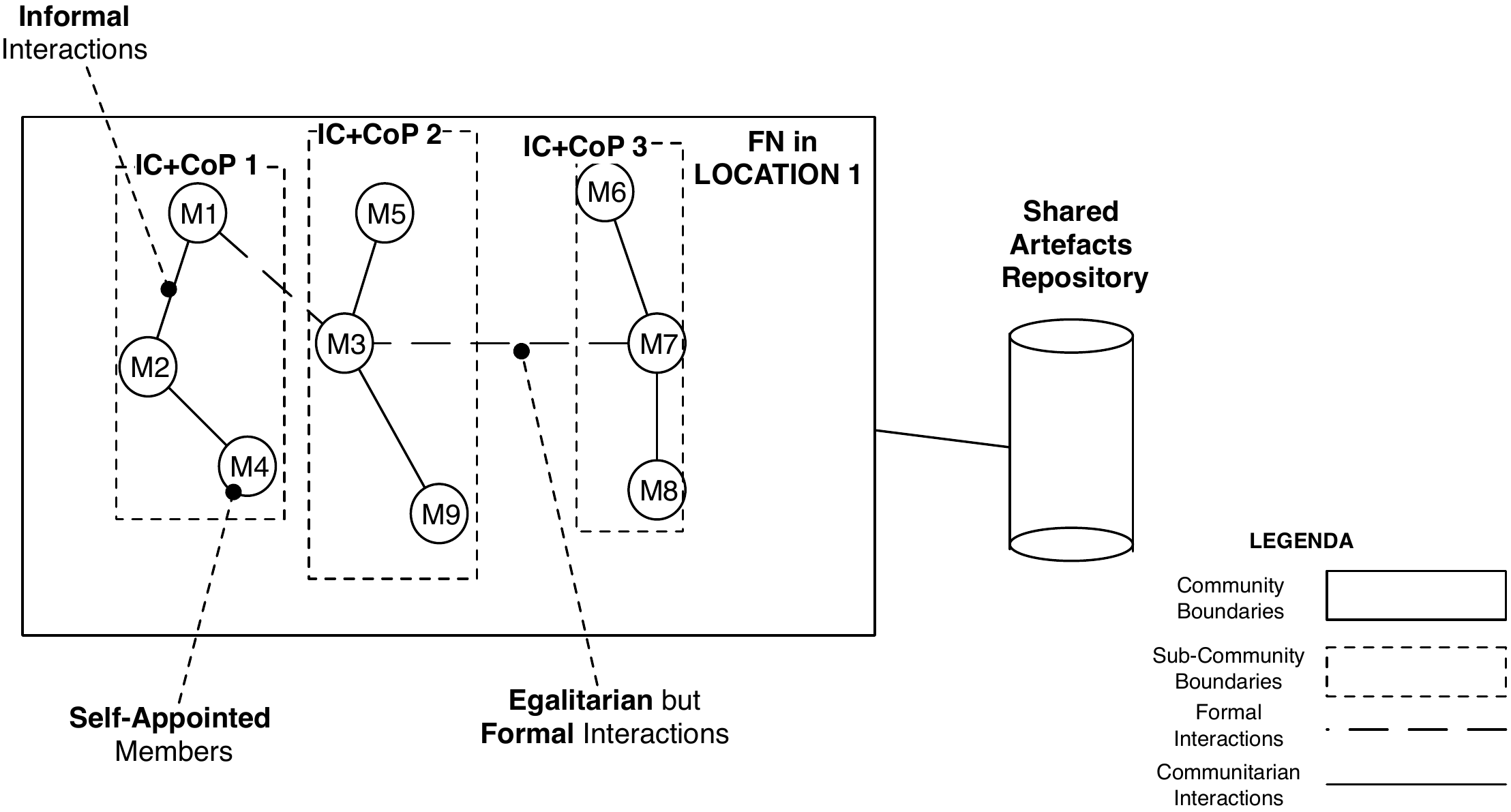}
\caption{Pattern 1: Formally-Networked Informal CoPs; note that Mx indicates a member.}\label{p1}
\end{figure*}

\textbf{Description and Operation.} This pattern occurs twice in our dataset: projects 1 and 2.  A software community reflecting this pattern would appear as in Fig. \ref{p1}. The pattern  reflects a \emph{co-located} set (i.e., with member distance $<30m$, LOCATION 1 in Fig. \ref{p1}) of internally-informal communities of practice (i.e., IC+CoPs, see dotted boxes in Fig. \ref{p1}) which is also formally-networked as part of a larger formal organisation. For example, think of small sets of self-appointed practitioners working in a practice- and role-based fashion around single software components (modules, features, etc) but also spread across that single location, while regulated by formal interaction agreements and sharing a regimented repository typical in CoPs and FNs.

\textbf{Interpretation and Discussion.} Because CoPs are by definition co-located and do not function outside of that location \cite{Wenger:1998,RamchandP12,ossslr}, this organisational structure pattern exhibits two distinct operational phases: (1) in one phase, the organisational structure works as an informal community of practice (IC+CoP) alone, no FN interactions occur; (2) in a second phase, the organisational structure works as a formal network organised around established interaction protocols (IC+CoPs+FN), for example when the integration of locally-constructed components requires interactions with peers outside the CoP.  This second organisational phase occurs when needed interactions exceed immediate proximity (i.e., $>30m$ \cite{Prikladnicki12,BjarnasonSER16}). This is because a formal network is distributed in nature, by definition \cite{ossslr}. In other cases when proximity is no longer an issue, communication and interactions are communitarian as part of an informal community, for example when the small CoPs switch back to refactoring their own software component and require increased informal interactions. 

\subsubsection{P2: Formally-Networked, Distributed Informal Communities (IN,IC,FN)} 
\begin{figure*}
\centering
\includegraphics[scale=0.6]{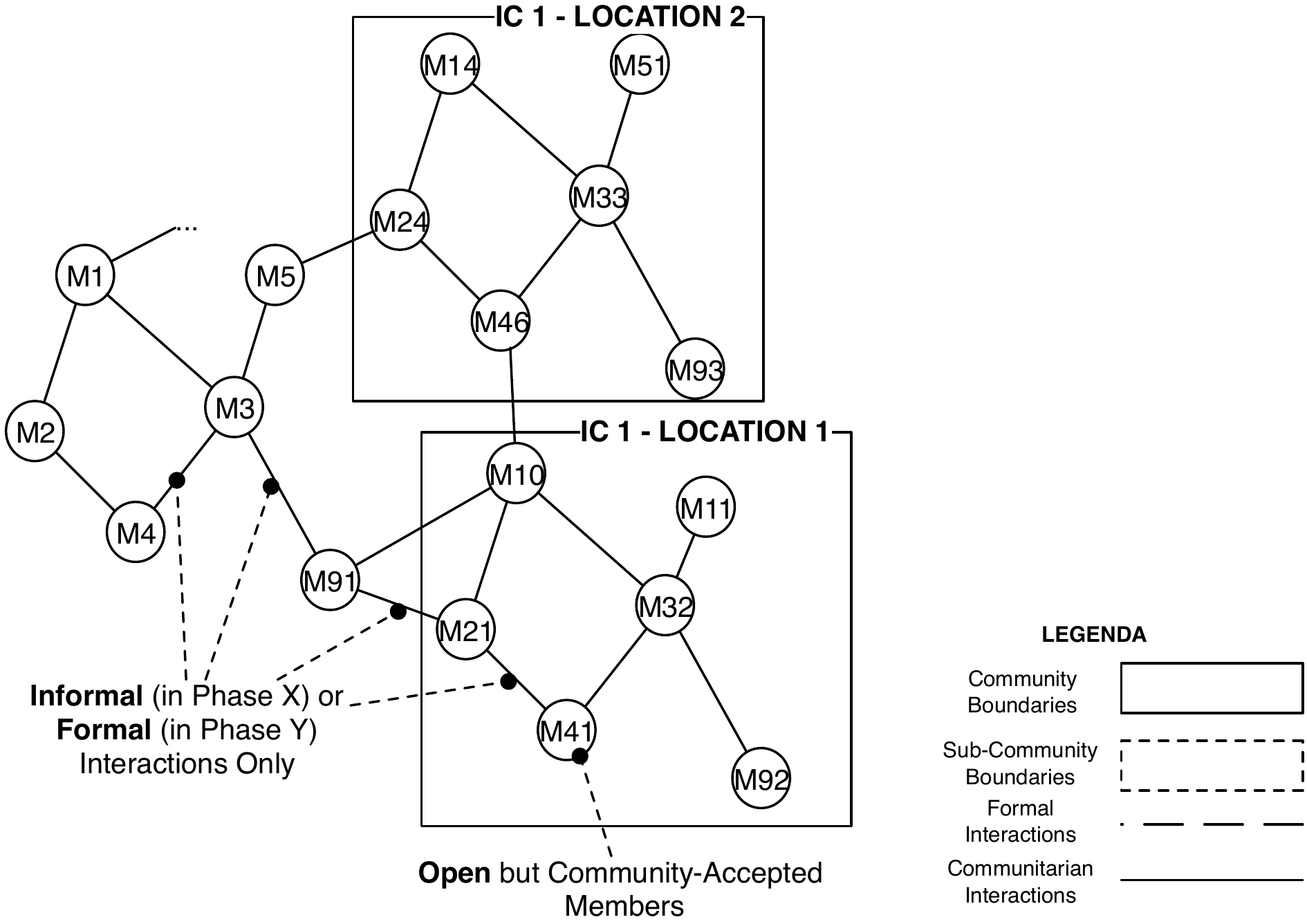}
\caption{Pattern 2 - Formally-Networked, Distributed Informal Communities; note that Mx indicates a member..}\label{p2}
\end{figure*}

\textbf{Description and Operation.} This pattern occurs twice in our dataset: projects 13 and 15.  A software community reflecting this pattern would appear as in Fig. \ref{p2}. The figure  shows distributed informal communities (i.e., ICs, see dotted boxes in Fig. \ref{p2}) which are also formally-networked (i.e., FNs, see square box in Fig. \ref{p2}). 

\textbf{Interpretation and Discussion.} Since the FN and IN types are mutually exclusive \cite{ossslr}, this organisational structure pattern exhibits two operational phases: (1) in one phase, informal communities are networked and share information and interactions as part of an Informal Network (IN+IC). For example, think of an open-source community composed of closely collaborating friends who are globally distributed; (2) in a second phase, the informal communities are networked and share information as part of a formal network (IC+FN).  For example, think of a Scrum-of-Scrums instance where formal protocols or agreements steer global collaboration.

\subsubsection{P3: Formally-Networked, Informal Project Teams (FN,IC,PT)}
\begin{figure*}
\centering
\includegraphics[scale=0.56]{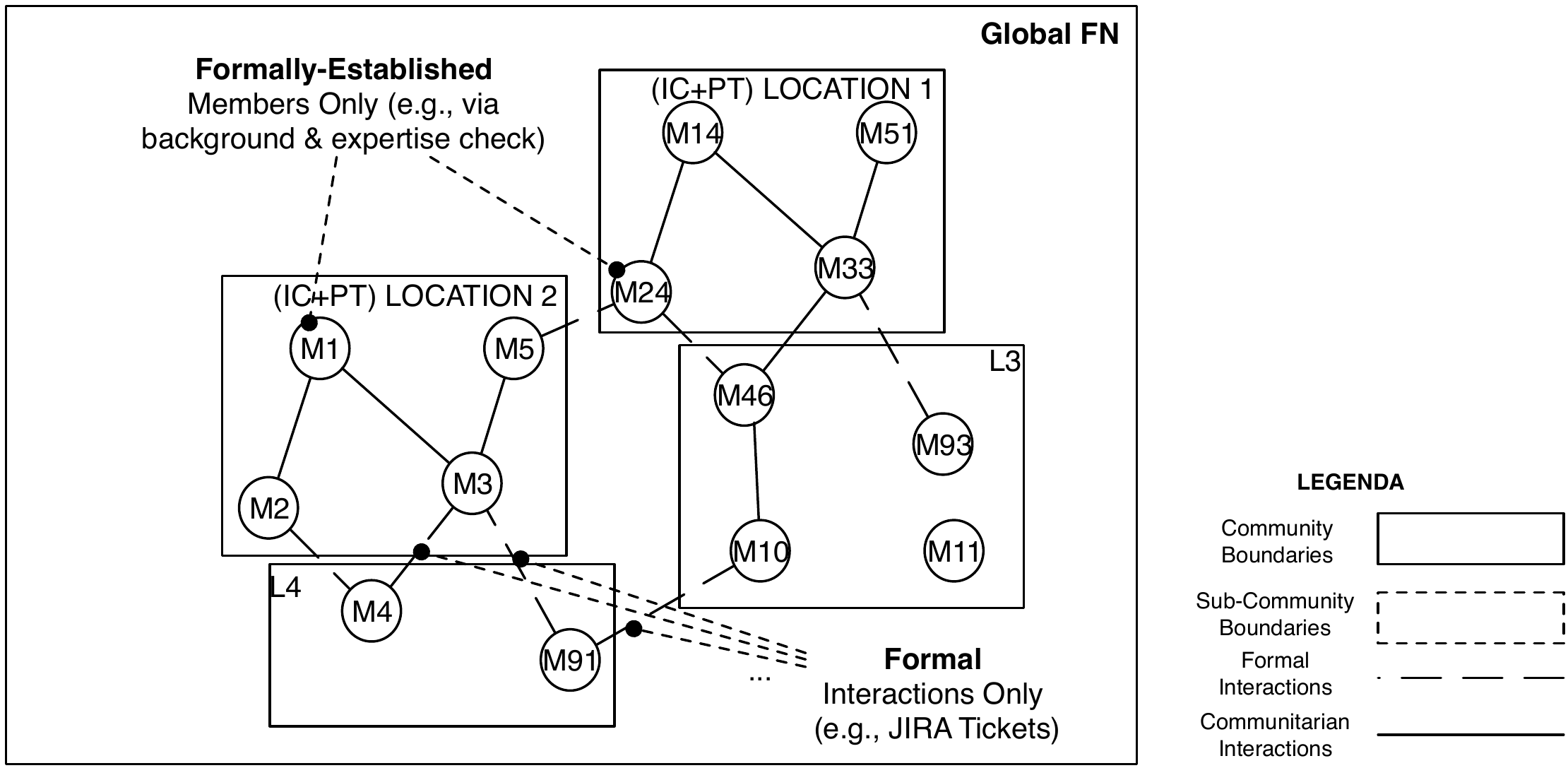}
\caption{Pattern 3 - Formally-networked informal community of project teams; note that Mx indicates a member.}\label{p3}
\end{figure*}

\textbf{Description and Operation.} This pattern occurs twice in our dataset: projects 3 and 27.  A software community reflecting this pattern would appear as in Fig. \ref{p3}. The figure  shows co-located project teams informally collaborating internally (IC+PT) but formally networked and spread across a wider global formal network. 

\textbf{Interpretation and Discussion.} This organisational structure pattern also occurs in closed-source global software engineering \cite{rel2} where small project teams essentially operate as small, informal co-located communities, often interacting well beyond the boundaries of their respective organisations, but interconnected to other distant project teams across a formally-specified and regulated formal network. For example, organisations which heavily rely on outsourcing fall into this pattern. This pattern is also found in well-structured and governed open-source initiatives and large ecosystems such as the Apache software foundation. In these relatively regimented open-source contexts, formal interaction protocols are upheld as part of membership agreements within the Apache software foundation forge (i.e., the FN), while smaller project-communities internally operate more or less as distributed project teams.

\subsubsection{P4: Informally-Networked, Situated, Informal Working Groups (CoP,IN,IC,WG)}
\begin{figure*}
\centering
\includegraphics[scale=0.5]{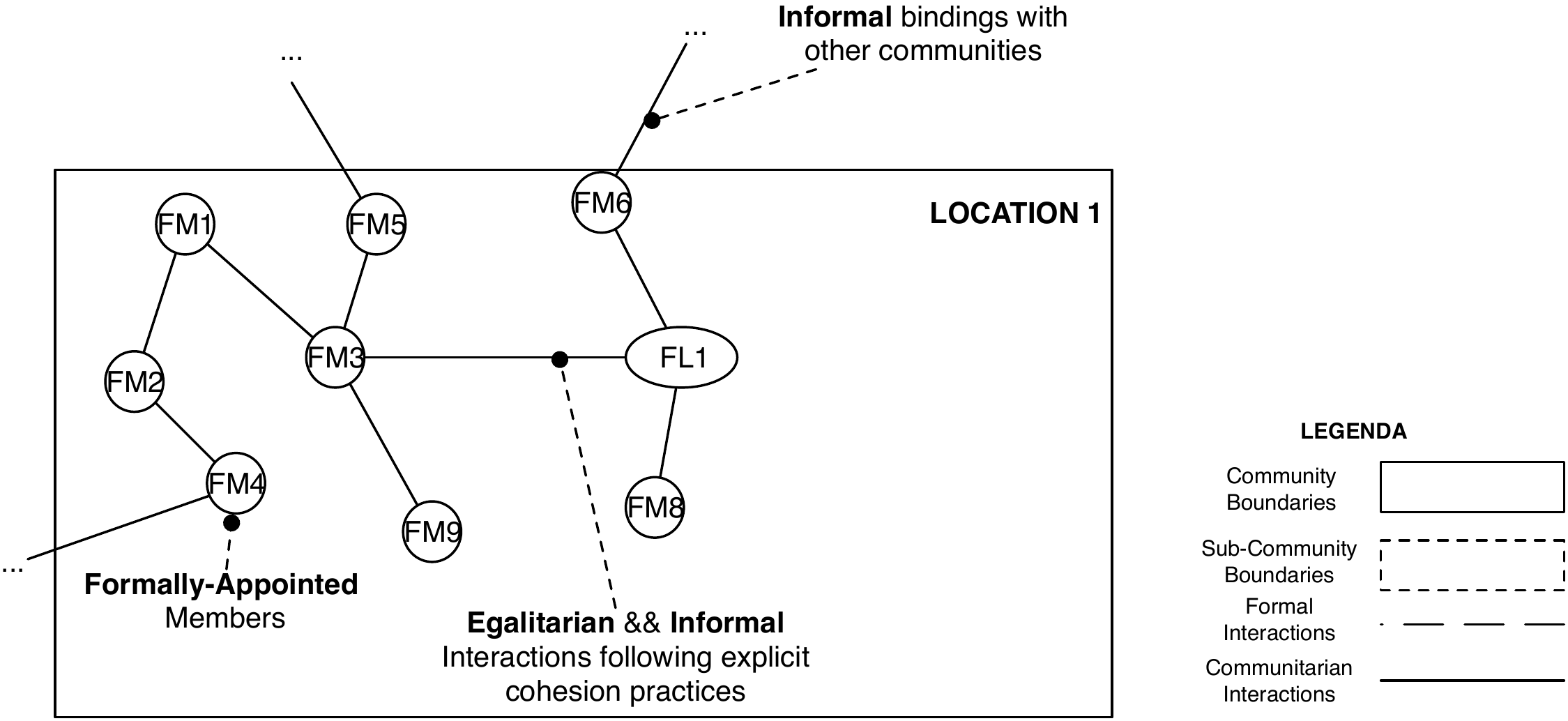}
\caption{Pattern 4 - Informally-Networked, Co-Located, Informal Working Groups; note that Mx indicates a member.}\label{p4}
\end{figure*}

\textbf{Description and Operation.} This is the most common organisational structure pattern in our dataset, occurring 9 times, in project IDs 1,2,7,8,9,13,18,21 and 22. A software community reflecting this pattern would appear as in Fig. \ref{p4}. The four primary identifiers for the types in the pattern are:
\begin{enumerate}
\item \textbf{Informal Interaction}
\item \textbf{Situatedness}
\item \textbf{Tight Cohesion}
\item \textbf{Personal Engagement}
\end{enumerate}
Consequently, this pattern reflects tightly-knit, cohesive, small, informal and co-located groups of practitioners (CoP+IC+WG) who are not practicing a specific role but rather contribute to addressing a common goal (as defined in WGs). As part of this goal, the tight group also networks informally with the outside world, participating in communities, as needed, address their goal (IN).
 
\textbf{Interpretation and Discussion.} Looking at the primary characteristics that identify this pattern, we were not surprised for one reason: the 4 characteristics highlighted above clearly reflect the main organisational goals behind agile methodologies, \emph{which are an underlying assumption of this study}. On one hand, this validates the analysis process and results reported in this article but, on the other hand, this finding introduces a construct validity threat into our study, since the results and discussions we have reported may have been tainted by the nature of our study subjects.  This and other threats to validity are fully discussed in Sec. \ref{threats}. Notwithstanding, this pattern shows a number of interesting properties.  For example, our quantitative analysis reveals that this pattern reflects the ``youngest" teams in our dataset, rather than more established ones suggesting that this patter is a sort of entry-point, rather than a stable organisational form. Moreover, this pattern reflects the highest number of architecture issues we reported; this suggests that the pattern is not a ``silver-bullet" for agile methods but actually the opposite. These and other quantitative reflections over the patterns found are available in Sec. \ref{quant}.

\subsubsection{P5: Informally-Networked, Informal Working-Groups (IN,IC,WG)}
\textbf{Description and Operation.} This pattern occurs 3 times and is the distributed variant of the CoP,IN,IC,WG pattern. These patterns differ on the lack of situated action \cite{situatedness}. With reference to Fig. \ref{p4}, this pattern variant adds a series of geographically distinct locations, distributing members across those locations and containing them into tightly knit informal working groups. 

\textbf{Interpretation and Discussion.} This pattern is consistent with the \emph{Scrum-of-Scrums} \cite{MaranzatoNH11,PaasivaaraLH12} variant of the pattern in Fig. \ref{p4}. For example, imagine a series of geographically distributed agile teams that cooperate informally towards their prescribed global goal.

\begin{figure*}
\centering
\includegraphics[scale=0.45]{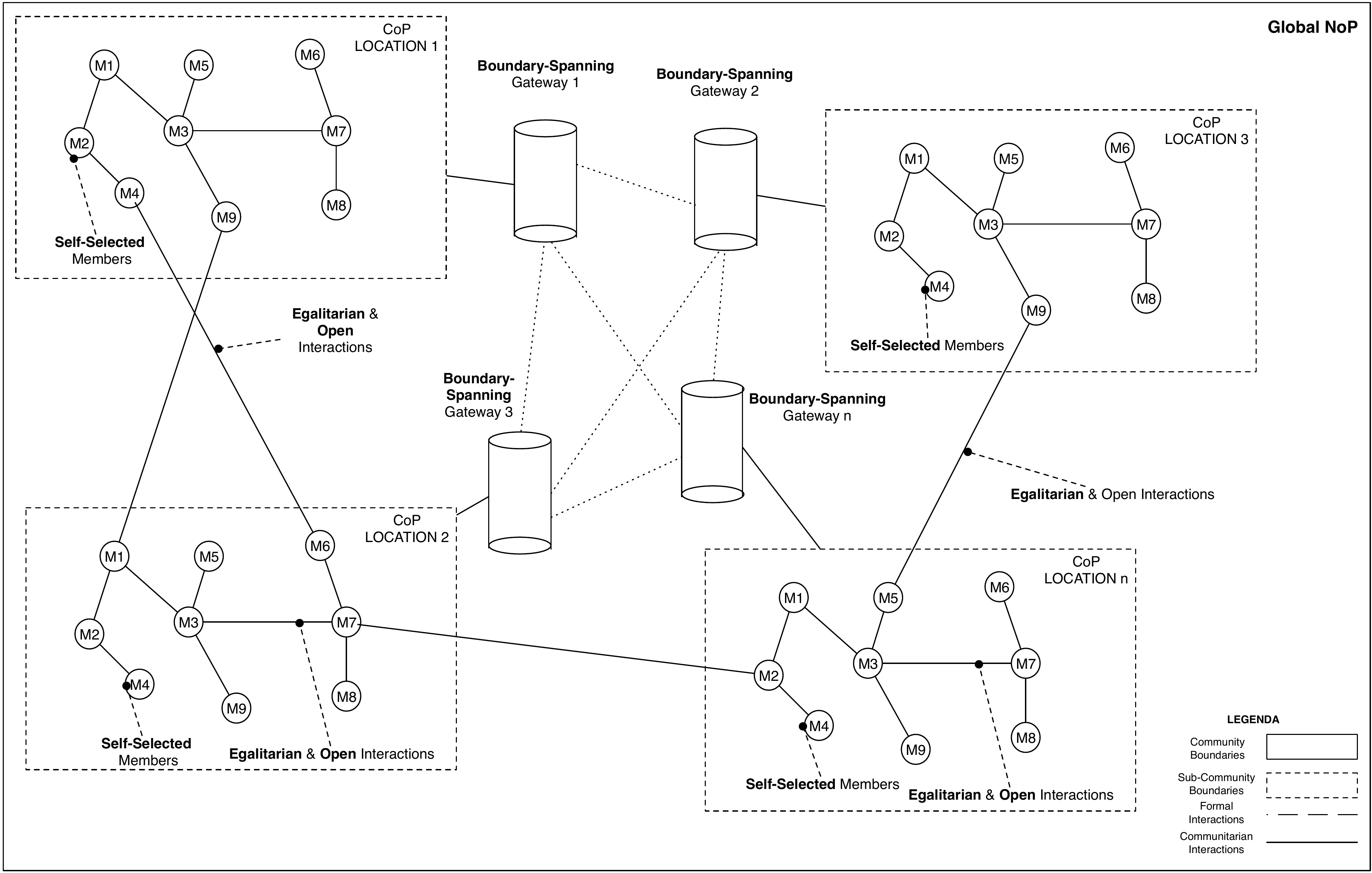}
\caption{Networked, Informal Community of Practices.}\label{q5}
\end{figure*}

\subsubsection{P6: Problem-Solving, Situated, Informal Working Groups (CoP,IC,WG,PSC)}
\textbf{Description and Operation.} This pattern occurs 3 times and is another variant of the pattern in Fig. \ref{p4}, differing solely by the prescribed goal of the working group being structured around a well-defined and clearly measurable problem (e.g., a technical issue such as software integration or testing). 

\textbf{Interpretation and Discussion.} With reference to Fig. \ref{p4}, this pattern variant would add a specific role to each node in the community, with a characterisation of that role with respect to the problem to be solved. For example, think of a tightly knit, situated informal group working on software quality improvement: several members are dedicated to analysis and modelling while others might prepare and execute test-cases.

\subsubsection{P7: Networked, Informal Community of Practice (CoP,IN,NoP)}

\textbf{Description and Operation.} This pattern occurs 3 times and reflects an organisational structure such as the one depicted in Fig. \ref{q5} where several situated communities of practice cooperate informally across global distances by means of shared repositories of knowledge, artefacts, and assets. 

\textbf{Interpretation and Discussion.} This pattern is the default mode of operation in global software engineering; its occurrence confirms results in several state-of-the-art reports from that field \cite{icgseoss,commflow,valbook}.

\subsection{Quantitative Analysis: Addressing Study Conjectures}\label{quant}

From a {\em quantitative} perspective our data conferred many surprising findings with respect to our initial conjectures.

\subsubsection{Organisational Characteristics that Mediate Architecture Quality}

From Table \ref{chars} we report that two characteristics are moderately correlated to occurring software architecture issues, namely, \emph{Cohesion} and \emph{Culture-Tracking}. While correlation does not mean causation, we suggest plausible explanations for these results. The first ---cohesion---suggests that a tighter organisational structure reflects a lower occurrence of architecture issues. This may be explained by the increased flow of architecture knowledge in a tighter organisation. 
What we found surprising is the relatively low correlation of the cohesion factor when compared to the presence of culture-tracking mechanisms, including architecture knowledge management \cite{WeinreichG16} devices (e.g., semantic wikis for architecture knowledge \cite{BoerV11}). The stronger correlation between the presence of culture-tracking mechanisms and the occurrence of architecture issues suggests that managing explicit architecture knowledge may be more beneficial than managing teams (and team cohesion) more effectively. 
We also note that team cohesion can be tracked and monitored whereas the effect of adopting software architecture knowledge management mechanisms may only manifest benefits on the long run, if at all.

\subsubsection{Addressing Conjecture 1: Characterising the Empirical Relation Between Architecture and Organization}\label{relation}

Our initial conjecture was: if there is a recurrent organisational structure pattern, then it may influence the quality of a software architecture, by virtue of the relation posited by Colfer and Baldwin's ``Mirroring Hypothesis'' \cite{colfer2010mirroring}. 

\begin{figure}
\centering
\includegraphics[scale=0.47]{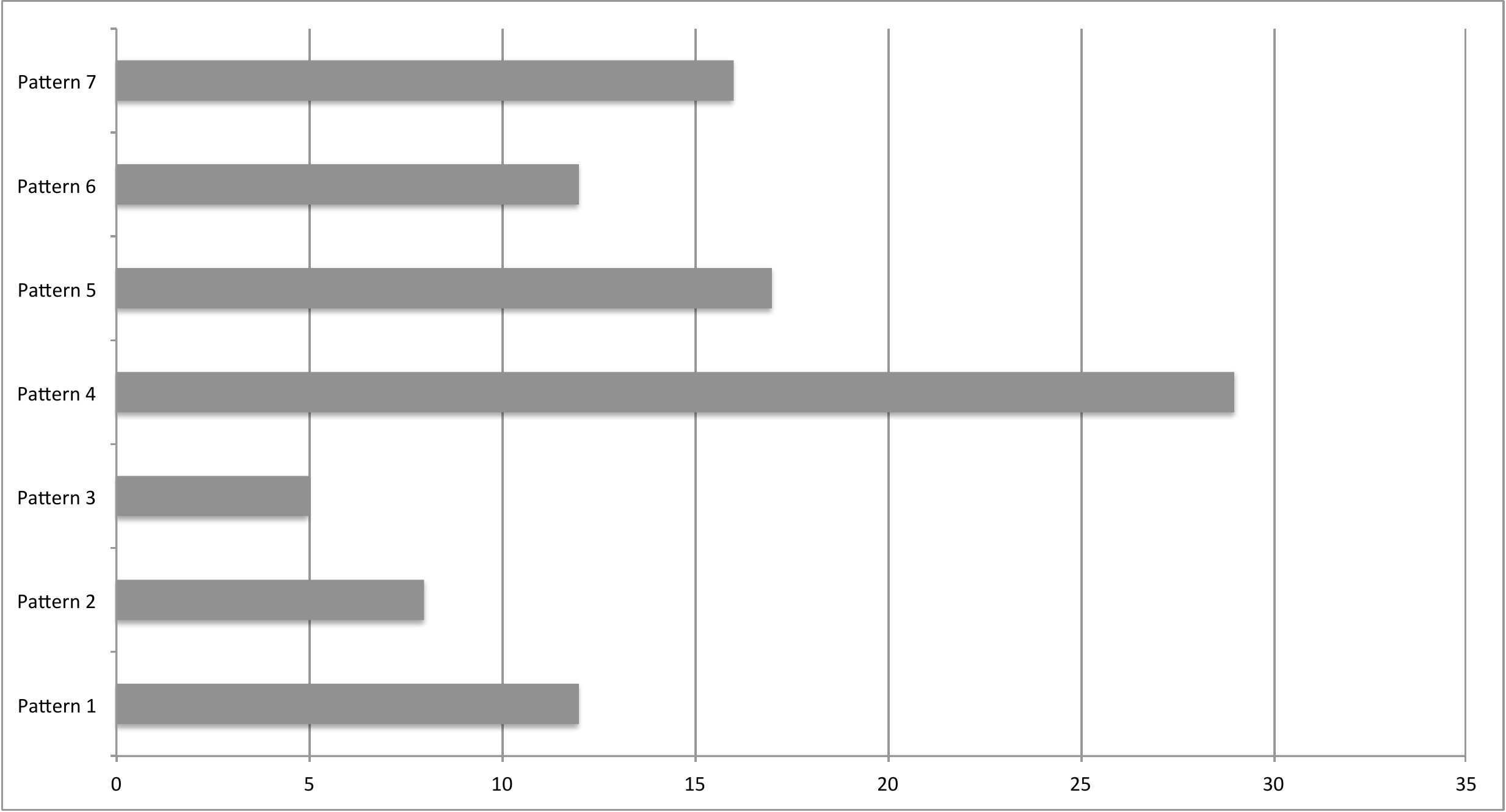}
\caption{Architecture Debt: An Overview for Reported Patterns. The Y-axis reports the patterns while the X-axis reports the total number of smells associated with all organisations exhibiting that pattern; Fig \ref{normsmells} plots the same numbers but with normalised \# smells with respect to \# organisation.}\label{patterneffic}
\end{figure}

To quantitatively assess this conjecture, we measured the architecture {\em debt} of the 7 patterns we found. (Recall from Sec. \ref{bg} that architecture debt of a pattern is defined as the count of architectural flaws and issues corresponding to that pattern.)  The raw counts are plotted in Fig. \ref{patterneffic}. 
\begin{framed}
\textbf{Finding 2.} In line with what we conjectured, a single most recurrent pattern does exist---pattern 4, occurring 9 times---and it corresponds to the highest architecture \emph{debt} in our sample. The same trend is evident also normalising the involved quantities. 
\end{framed}

This is consistent with---and lends empirical support to---Colfer and Baldwin's Mirroring Hypothesis \cite{colfer2010mirroring}. 
Our rationale for this finding is that Pattern 4 is the most ``informal" of the patterns, and so the social network of participants is the most highly connected \cite{crowston06}, and hence the architecture mirroring this is also highly connected.  But highly coupled architectures are hard to maintain and evolve, hence they have the most smells.
Another consequence of that hypothesis within our study, by the same logic, is that the remaining patterns should be associated with better architectures, i.e., fewer flaws and hence lower architecture debt. 

\begin{figure}
\centering
\includegraphics[scale=0.44]{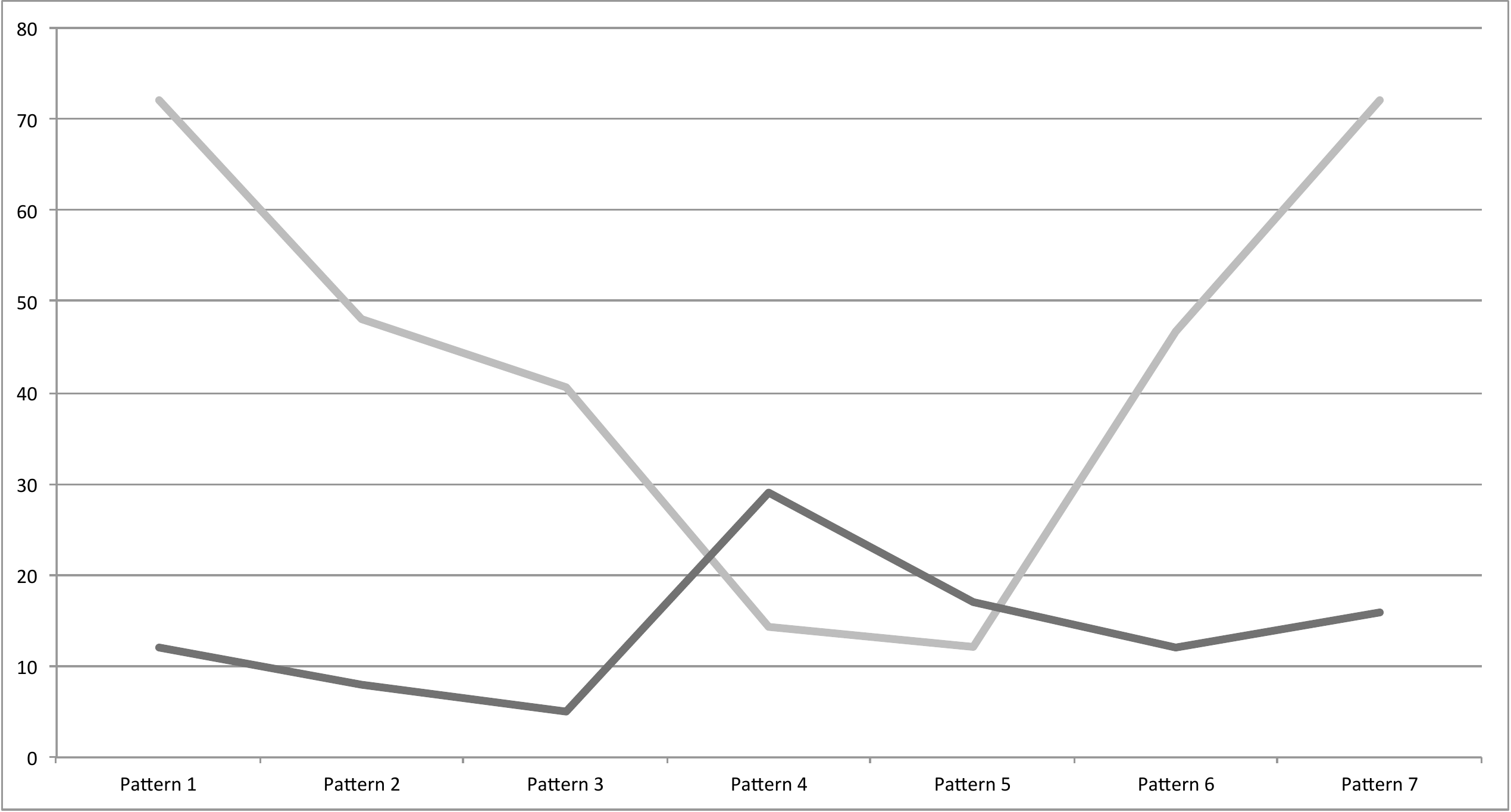}
\caption{Architecture debt (darker curve in the bottom-half) vs. Team age (lighter curve in the top half).}\label{archeffic}
\end{figure}

\begin{framed}
\textbf{Finding 3.} 
Furthermore, we reported an even higher negative correlation of -0.48 between architecture debt of reported patterns and their team age, as highlighted in Fig. \ref{archeffic}. This finding is intuitive and aligns with earlier research \cite{xiao:2016} suggesting that, as teams and organisations mature, their software architecture debt decreases. 
\end{framed}

According to our data, organisational structures need to ``mature", to reach organisational stability, at which point the structure and its underlying architecture stabilise. On the other hand, according to our empirical evidence, community \emph{size} is not significant---in particular, correlation values between community size and architecture issues are negligible, at around 0,2.

\subsubsection{Addressing Conjecture 2: An organisational evolution model for software organisational structures}\label{evol}

As previously introduced, our second conjecture was that: ``if an organisational structure pattern can be found, then it reflects `veteran' teams, according to the principle of organisational stability, well-established in organisations research and social networks analysis \cite{QuintanePRM13,WangK01}".

\begin{figure}
\centering
\includegraphics[scale=0.44]{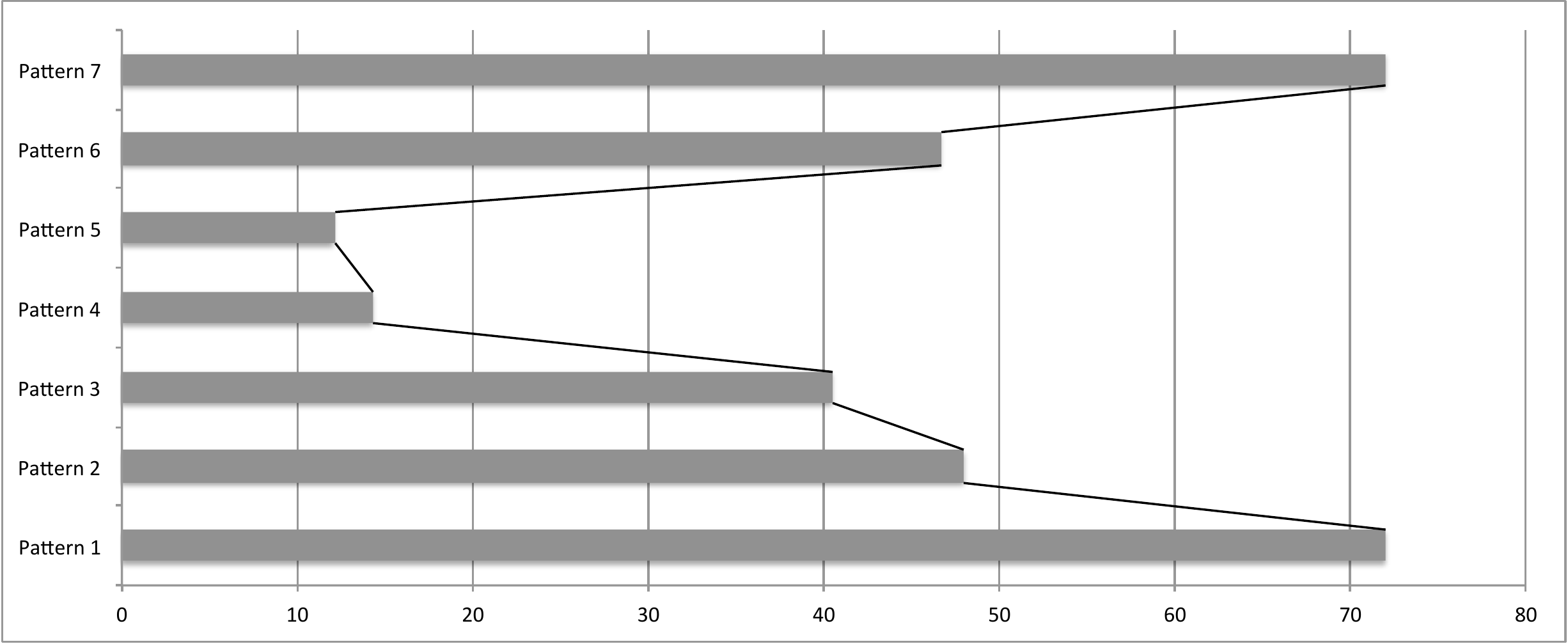}
\caption{Team age (X-axis, in months) trendlines on Ordered Patterns by increasing formality (Y-axis, from bottom to top) based on the presence of more formal organisational structures; team ages are averaged across the organisations of each pattern;}\label{trendlines}
\end{figure}

To further understand the role of team age with respect to the organisational structure patterns found, we plotted trend lines for reported team ages associated with each pattern, and ordered the patterns in terms of increasing informality (see Fig. \ref{trendlines}).  The figure shows team age (X-axis, in months) trendlines plotted over the occurring patterns ordered by increasing *formality* (Y-axis, from bottom to top)  and where team ages are averaged across the organisations of each pattern.

\begin{framed}
\textbf{Finding 4.} 
Our data shows that the most frequent patterns occur in correspondence with the youngest teams (P4 and P5), and sits exactly at the middle of the plot in Fig. \ref{trendlines}. This means that the patterns in question are a point of balance between \emph{informality} and \emph{formality}. Furthermore, two super-linear trend lines can be plotted outgoing from that central point: as team age increases from that point the patterns diverge evenly towards informality and formality. 
\end{framed}

\begin{figure}
\centering
\includegraphics[scale=0.45]{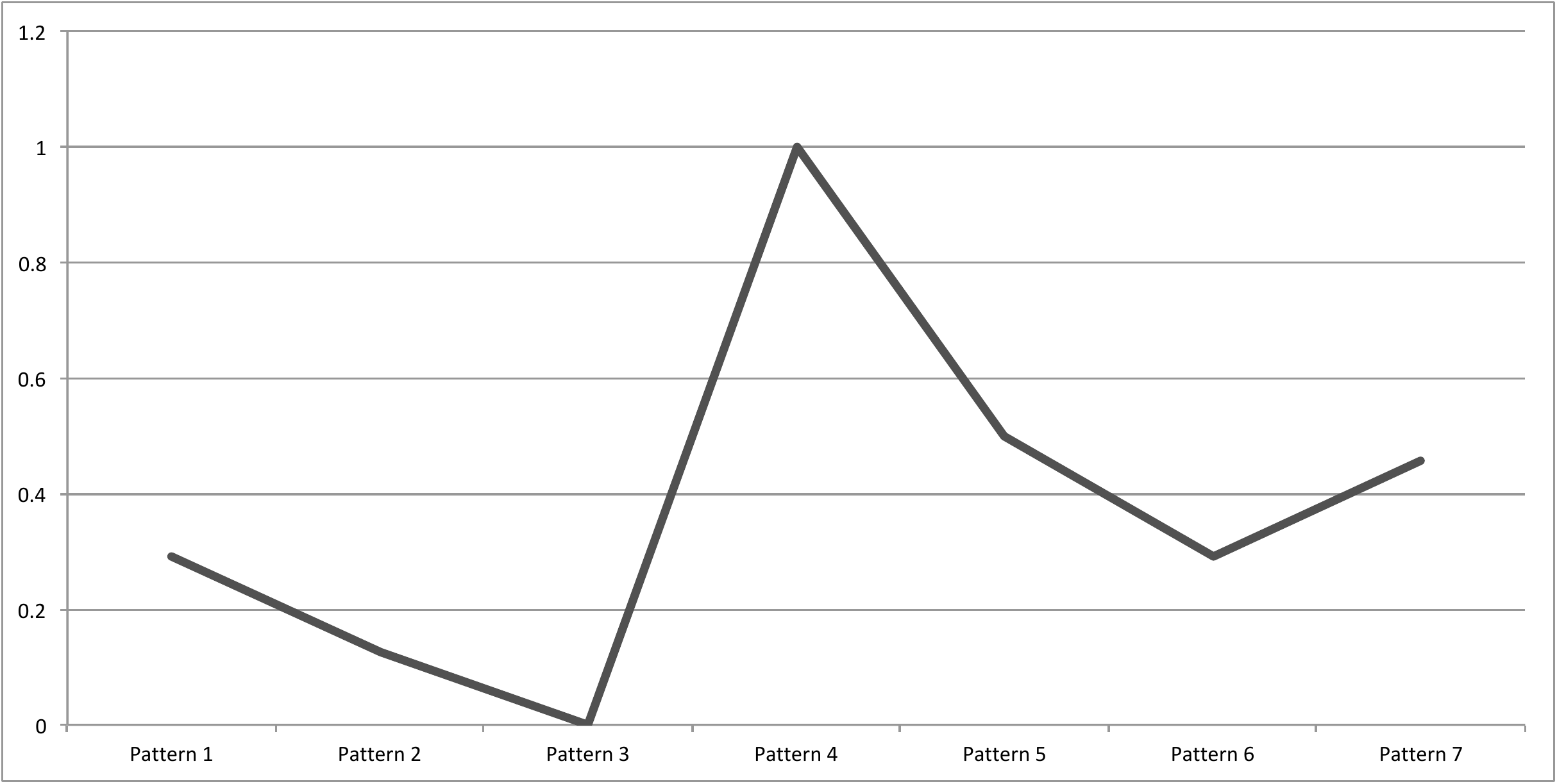}
\caption{Architecture debt: An Overview for Reported Patterns---the X-axis reports the patterns while the Y-axis reports the total number of smells normalised \# smells w.r.t. \# organisation.}\label{normsmells}
\end{figure}

\subsection{Findings Overview, Interpretations, and Outlook}

Table \ref{findings} synthesises our findings. Our interpretation is that the role of organisational structures for software architecture design and their quality maintenance is loose at best, limited to two (cohesion, and culture-tracking) out of thirteen possible characteristics in our study alone. 
Although this  interpretation may be due to a sampling issue (our sample was constituted by agile teams only) the evidence clearly indicates that a higher interoperation between organisational structure patterns (via their measurable characteristics) and software architectures may yield better general quality of software architecture, e.g., as reflected by a lower number of architecture issues. For example, Finding 1 and 2 confirm that organisational structure patterns do exist, which indicates importance of the issue, while Findings 3 and 4, reflect a moderate relation between patterns and architecture quality. Conversely, Finding 5 confirms that several previously-observed characteristics (formality/informality) in our case may be mediating heavily in organisational structure and software architecture quality. From these findings, \emph{we conclude that organisational structure quality measurement and improvement is important, and should be further studied, not only for organisational stability, quality enrichment using insights from organisations and management research \cite{jisaspecissue}, but also for stability measurement \cite{Lehman1980} and general software architecture quality itself.}

Our evidence and findings also indicate that an alignment may manifest between the two structures, \emph{eventually}, and with increasing team age. Findings 3 and 5 indicate that if such an alignment is achievable, it would manifest beyond our 72-month analysis time window.  Also, the same findings  suggest that such an alignment currently can only manifest \emph{*by chance*}, rather than \emph{*by design*}, since none of the organisations we studied report an engaged and active usage of organisational structure models or measurements to drive software architecture quality, or architecture evolution. \emph{We therefore propose that organisational structure models and their role for software architecture design should be investigated further from a practical and theoretical perspective; their usage may prove vital to improve software architecture and software community quality}.

What is more, the alignment in question seems to reflect two distinct ways of organisational working, one formal, and one informal, with different characteristics that may be positively or negatively affecting software architecture quality. This suggests that Crowston \& Howison's conjecture about informality being a trigger for open-source community quality \cite{Crowston03thesocial} is valid for closed-source projects as well. But our evidence also reflects the importance of the opposite end of the spectrum---{\em formality}---which is equally important. Also, this duality is curiously reminiscent of Kahneman's theory of System-1 (thought-processes are more informal, irrational, free-thinking, associative) and System-2 thinking (thought-processes are more structured, rigorous, formal) \cite{Kahneman2012}, whose applicability to software design has been conjectured several times\footnote{\url{https://www.targetprocess.com/blog/2013/07/software-development-fast-and-slow/}},\footnote{\url{https://resources.sei.cmu.edu/asset_files/Presentation/2015_017_101_438769.pdf}} but never  proven. Hence further investigation of the influence of organisational structure metrics and design decision-making processes could reveal software engineering behaviour patterns which are more (or less) efficient for specific software engineering roles such as that of software designers.

\begin{table}
\caption{Summary of Findings.}\label{findings}
\begin{tabular}{|c|>{\raggedright}p{10.5cm}|}
\hline 
\textbf{ID} & \textbf{Finding}\tabularnewline
\hline 
{\footnotesize{}1} & {\footnotesize{}7 non-overlapping patterns, consisting of at least 3 community
types, recur across our dataset and a single pattern
covers more than 37\% of our dataset.}\tabularnewline
\hline 
{\footnotesize{}2} & {\footnotesize{}The single most recurrent pattern corresponds to the highest
architecture \emph{debt} in our sample.}\tabularnewline
\hline 
{\footnotesize{}3} & {\footnotesize{}A higher negative correlation of -0.48 between architecture
debt and team age indicates that as teams and organisations
mature, architecture efficiency increases. }\tabularnewline
\hline 
{\footnotesize{}4} & {\footnotesize{}The most frequent pattern occurs on the youngest teams, and
represents an entry-point at middle of Formality/Informality plots
- the pattern in question is a point of balance between \emph{informality}
and \emph{formality}.}\tabularnewline
\hline 
\end{tabular}
\end{table}

\subsection{Limitations and Threats to Validity}\label{threats}

Like any study of comparable magnitude and scale, this study is affected by several threats to validity \cite{wohlin}. In what follows we outline the major ones in our study design and execution.

\subsubsection{Internal and Sampling Validity}

Internal validity refers to the internal consistency and structural integrity of the empirical research design.  Specifically it refers to how many confounding factors may have been overlooked. For example, the patterns we report cover 24 out of 30 datapoints, with 6 outliers that were ultimately discarded. This issue may negatively affect the generalisability of our results since those datapoints may reflect unknown community types or characteristics and quantities that emerge with a more fine-grained analysis lens (e.g., focusing on single software artefacts and the communication/collaboration around them). We attempted to address this threat with a sampling strategy where we controlled as many variables as possible to: (a) ensure a meaningful variety of our sample; (b) ensure that important variables for the objects of our study, namely organisational structure and architecture, were controlled. For example, we controlled for organisational maturity influences over organisational structures by selecting agile teams that had recently successfully adopted agile methods. Also, we controlled for process maturity, by selecting a heterogeneous sample according to a standard CMMI scale (although we acknowledge that this scale is not fully consistent with current agile practices and teams). Furthermore, there are up to 90 factors from the state of the art in organisation research \cite{ossslr} that may  be affecting our findings and results; also, the quantities and effect size of the factors themselves were not addressed in this study at all. Stemming from this limitation, we are planning an additional study of our target subjects in follow-up research that will further address the validity of this work.  

Furthermore, the statistical validity of the results reported in this manuscript relies heavily on multiple comparisons among possibly related quantities, which leads to a statistical analysis issue known as the ``Multiple Comparisons Problem" \cite{gelman2013garden}. This condition is typically addressed by increasing sample size and correcting with the well-known (but overly-conservative) Bonferroni correction \cite{TeradaS13}.  Because our study was exploratory, we did not perform oversampling or apply the afore-mentioned correction so our findings remain potentially affected by this issue. In the future we intend to replicate this work to further generalise the findings and increase their validity. In this new study we are planning to design the analyses for hypothesis testing, factoring in the possible control operations entailed by tests such as the Bonferroni or FDR correction.

\subsubsection{Construct and External Validity}

As previously discussed, our decision to strengthen internal and sampling validity by focusing on agile teams alone inherently introduced a flaw in our construct validity.  Also, although our measurements, observations, and findings are based on valid content (i.e., reported by practitioners who were directly involved with and witnesses to the reported subjects) and valid criteria, the external validity connected to the above-mentioned flaw may be compromised as well. For example, we used simple non-weighted and aggregate sums to evaluate the quantities involved in this study so we have no way of knowing whether the entity and \emph{-arity} of the involved quantities may yield different results.  We are planning further studies using a more quantitative approach including rigorous statistical modelling and testing. We have gained confidence that our findings and observations are valid, at least for agile teams, which currently\footnote{\url{https://techbeacon.com/survey-agile-new-norm}}\footnote{\url{http://stateofagile.versionone.com/}} account for 76\% of modern software engineering teams. However, as remarked for internal and sampling validity as well, our study design is flattened with respect to several of the quantities involved. For example, architecture issues were not counted but rather our triangulated Delphi study simply confirmed their presence as a 0-1 coefficient. Hence, it could  be that another organisation would have one most prominent and recurring smell in project A and two less prominent, less recurring smells in another project B. In such a case, using our current measure, the following relation would hold:

\begin{center} 
ArchDebt(B) $>$ ArchDebt(A);
\end{center}

But, as yet, we have no way of knowing or generalising whether this is true or not beyond the samples and scope of our study. What is more, the two quantities we correlate with our statistical analysis could be trivially correlated by construction, namely, the commonality of a pattern (the number of projects exhibiting that pattern) is trivially correlated (i.e., by construction) with Architectural debt: the more projects there are (of some pattern), the more architectural issues will be reported. Although this is true in principle, it is not {\em necessarily} the case. That is, this was not established in practice for the phenomena we are studying in this context. The statistical assumption behind a linear correlation between two co-increasing quantities such as the ones being discussed in this issue may not hold since the phenomena we are observing are \emph{macro-}phenomena of organisational and socio-technical nature, a realm where an exploratory approach needs to be employed to empirically establish even the most blatant variation.

\subsubsection{Conclusion Validity}

Conclusion validity represents the degree to which conclusions about the relationship among variables are reasonable. In the scope of the discussions of our results we made sure to minimise possible interpretations, designing the study with reference to known hypotheses. Also, our conclusions were drawn from statistical analysis of our dataset. This not withstanding, the conclusions drawn from our study also reflect assumptions which, although sound and reflecting the need to avoid research design mishaps, may compromise the conclusion validity.  For example, we chose to focus on the most basic roles for software practitioners in our sample, not reaching out explicitly, for example, to software architects. This decision may have compromised our ability to go beyond the simple numbers and capture speculative interpretations of architects who work on multiple projects, possibly within different organizations. Thus, our study remains subject to this threat.


\section{Related Work}\label{rel}

Perhaps the most relevant research related to our investigation comes from
Colfer et al. \cite{colfer2010mirroring}, who investigated the ``mirroring" hypothesis: that organisational structure (represented as a network of co-committing and communicating developers) and software architecture (represented as a design-structure matrix \cite{dsm}) are mirror images of each other. The authors found quantitative evidence in open-source which supports the mirroring hypothesis. However, they focused on a small number of projects and never actually looked at industrial settings may be different organisationally than open-source projects---these limitations greatly hamper the generalisability of their results.  A similar observation concerns the work reported by \cite{KwanCD12} and \cite{herbsleb1999beyondconwayslaw}. 

In the scope of this study, we strove to elicit the recurrent organisational structure patterns that take place in software engineering practice, by direct observation of the  ``trenches" in which those patterns operate. In the scope of our results, however, we also confirmed in industry the  same observations and conclusions that
Colfer et al. arrived at, namely, \emph{that there is a strong symbiotic mirroring relation between organisational structures and software architectures} and this relation goes beyond single factors such as socio-technical congruence \cite{CataldoHC08}, software developer truck factor \cite{TorchianoRM11} or similar valuable but overly narrow metrics.

In our study we set out to distill organisational characteristics that provide measurable structural quality evidence of organisations, by comparing with the technical structures (i.e., software architectures) on which they work \cite{cacmqualcomm}. By deepening our understanding over these organisational characteristics, we mean to provide measurement, prediction, and evaluation means through which, for example, both technical and social debt may be assessed and managed at once \cite{Brown10debt,Kazman15debt}.  In the same way, similar studies such as Howinson et al. \cite{alonetogether} have tried to distill a theory of socio-technical aspects in open-source (e.g., motivation, coordination or collaboration), but this and similar approaches \cite{MoonH14} fail to relate to concrete patterns (e.g., design patterns across a software architecture) and metrics (e.g., collaborativity or cohesion across an organisational structure) that can be used for planning corrective and preventive action. 

 From another related perspective, several works have addressed the role of organisational structures in distributed and global contexts (e.g., Lanubile et Al. \cite{LanubileEPV10,LanubileCE13}) as well as organisational alignment (e.g., Koch \cite{Koch92}) with software architectures and business objectives. Most prominently, in this area, figures the work by Betz et Al. \cite{BetzW12} who systematically investigate the practices and theories around Business, Architecture, People, Organizations.

\section{Progress Beyond the State of the Art}\label{sota}

\subsection{Software Engineering Research}
With respect to the previous work illustrated above, our work is a natural extension: our intents  to understand what patterns exist in organisational structures is reflecting on the newly-introduced notion of social communities in software engineering.  The patterns we uncover may provide a basis for further software organisation research, e.g., in the contexts of more effective agile or DevOps migration. Similarly, further research on the communities included in the patterns and how these patterns may be measured should also ensue. At the same time, further research needs to establish quantitatively how the patterns relate to (or align with) software architecture quality. As a first key step in this direction, for example, our findings confirm the previously unproven conjecture by Betz et al. who theorize that decisions in one dimension of the BAPO model influences the others. Further research into this interplay is needed to qualify and quantify  the concepts involved and their dimensions into controllable factors.

\subsection{Software Engineering Practice}

From a more practical perspective, our contributions can be used as follows.
First, the recurrent organisational characteristics reported in Sec. 4.2 can be used as a basis for organisational rewiring; measuring such characteristics practitioners can plan and drive their organisational rewiring exercises with instrumented decisions.
Second, the organisational patterns reported in Sec. 4.3 and fleshed out in Sec. 5 can be used as reference targets (or \emph{pitfalls} to be avoided) depending on \emph{fitness-for-purpose}.
Similarly, the issues and findings reported in Sec. 4.4 and Sec. 5.2 may aid in understanding the potential consequences of certain organisational scenarios and decisions, e.g., to study and enact corrective actions along the dimensions and characteristics outlined in Sec. 4.2.
Finally, the model we fitted with our data and fleshed out in Sec. 5 can be used by both practitioners and researchers. Practitioners can use it to plan organisational rewiring around two essential and mutually-exclusive characteristics namely \emph{formality} and \emph{informality}. These characteristics predominantly drive organisational rewiring and can be used as a compass in such exercises. And researchers can better explore and quantify the proposed model, e.g.,  by creating a more quantitative and generalisable basis.

\section{Conclusion}\label{conc}
This article reports on findings from a systematic, wide-spectrum investigation of organisational structure patterns in modern software industrial practice.  30 teams from 9 organisations were involved to provide a significant corpus of data as part of our mixed-methods study. The insights and models provided in this paper are useful to manage and steer the relationships between organisational structures and software architectures.

\subsection{Research Questions}

In addressing our main research question, namely, \emph{``is there a recurrent organisational structure pattern in modern software engineering that can be used as a reference, and how does it reflect on software architectures?"} We found that there is, in fact, a recurrent organisational structure pattern in modern software engineering organisational structures but it does {\em not} reflect good quality software architectures.  Also, the pattern may indicate some sort of entry-point in what can be considered an organisational evolution model for agile software engineering organisational structures---this entry point is driven by at least two variables: (a) team age; (b) formality levels. We observed that these two variables play a key role in establishing the organisational direction of structure patterns reported across our data.

The results and findings in this paper may prove useful for practitioners when re-wiring their organisational structures, e.g., as part of DevOps adoption, agile method changes, or similar software organisational re-design.

\subsection{Future Work}

In the future we plan to strengthen the validity of our findings by running systematic {\em quantitative} studies over the variables and relations we have observed. We plan to further explore the organisational structure patterns observed in this study by providing a metrics framework to observe them in action at larger scales. In particular, we will try to understand the degree to which architecture smells correspond to specific organisational types, stemming from quantifiable measures of organisational types. Moreover, we plan to refine the analyses we reported in this study to capture some of the insights and gaps we reported. For example, we intend to address the following research questions: (a) to what extent do organisational characteristics influence the emergence of each architecture smell? (b) what organisational structure pattern is most prominent in open-source communities? Furthermore, we plan to instrument a longitudinal study,  on a restricted set of the organisations that we involved in this study, to assess the followups of the insights reported in this article. Finally, we plan to replicate this study in industrial settings, increasing sample size and variety, to understand how our results may be suffering from the construct validity flaw discussed in Sec. \ref{threats}.



\bibliographystyle{acm}
\bibliography{smells}

\begin{thebibliography}{10}

\bibitem{commflow}
{\sc 0002, M.~J., and Piattini, M.}
\newblock Problems and solutions in distributed software development: A
  systematic review.
\newblock In {\em SEAFOOD\/} (2008), K.~Berkling, M.~Joseph, B.~Meyer, and
  M.~Nordio, Eds., vol.~16 of {\em Lecture Notes in Business Information
  Processing}, Springer, pp.~107--125.

\bibitem{Abbas09}
{\sc Abbas, N.}
\newblock {\em Software quality and governance in agile software development.}
\newblock PhD thesis, University of Southampton, UK, 2009.
\newblock British Library, EThOS.

\bibitem{paperaranda2008}
{\sc Aranda, J., Easterbrook, S., and Wilson, G.}
\newblock Observations on conway's law in scientific computing.
\newblock 1st Workshop on Socio-Technical Congruence (STC), at the 30th
  International Conference on Software Engineering (ICSE'08), Leipzig, Germany,
  10 May 2008, May 2008.

\bibitem{Baham16}
{\sc Baham, C.}
\newblock The impact of organizational culture and structure on the
  routinization of agile software development methodologies.
\newblock In {\em AMCIS\/} (2016), Association for Information Systems.

\bibitem{Bass17}
{\sc Bass, L.}
\newblock The software architect and devops.
\newblock {\em IEEE Software 35}, 1 (2018), 8--10.

\bibitem{kazmanbook}
{\sc Bass, L., Clements, P., and Kazman, R.}
\newblock {\em Software Architecture in Practice}.
\newblock SEI series in software engineering. Addison-Wesley, 2012.

\bibitem{Bertran11}
{\sc Bertran, I.~M.}
\newblock Detecting architecturally-relevant code smells in evolving software
  systems.
\newblock In {\em ICSE\/} (2011), R.~N. Taylor, H.~C. Gall, and N.~Medvidovic,
  Eds., ACM, pp.~1090--1093.

\bibitem{BetzW12}
{\sc Betz, S., and Wohlin, C.}
\newblock Alignment of business, architecture, process, and organisation in a
  software development context.
\newblock In {\em ESEM\/} (2012), P.~Runeson, M.~H�st, E.~Mendes, A.~A.
  Andrews, and R.~Harrison, Eds., ACM, pp.~239--242.

\bibitem{BjarnasonSER16}
{\sc Bjarnason, E., Smolander, K., Engstr�m, E., and Runeson, P.}
\newblock A theory of distances in software engineering.
\newblock {\em Information \& Software Technology 70\/} (2016), 204--219.

\bibitem{bh98}
{\sc Bowman, I.~T., and Holt, R.}
\newblock Software architecture recovery using conway's law.
\newblock In {\em CASCON\/} (1998).

\bibitem{Brown10debt}
{\sc Brown, N., Cai, Y., Guo, Y., Kazman, R., Kim, M., Kruchten, P., Lim, E.,
  MacCormack, A., Nord, R., Ozkaya, I., Sangwan, R., Seaman, C., Sullivan, K.,
  and Zazworka, N.}
\newblock Managing technical debt in software-reliant systems.
\newblock In {\em FSE/SDP Workshop on the Future of Software Engineering
  Research at ACM SIGSOFT FSE-18\/} (2010).

\bibitem{congruence}
{\sc Cataldo, M., Herbsleb, J.~D., and Carley, K.~M.}
\newblock Socio-technical congruence: a framework for assessing the impact of
  technical and work dependencies on software development productivity.
\newblock In {\em ESEM '08: Proceedings of the Second ACM-IEEE international
  symposium on Empirical software engineering and measurement\/} (New York, NY,
  USA, 2008), ACM, pp.~2--11.

\bibitem{CataldoHC08}
{\sc Cataldo, M., Herbsleb, J.~D., and Carley, K.~M.}
\newblock Socio-technical congruence: a framework for assessing the impact of
  technical and work dependencies on software development productivity.
\newblock In {\em ESEM\/} (2008), H.~D. Rombach, S.~G. Elbaum, and J.~M�nch,
  Eds., ACM, pp.~2--11.

\bibitem{Chatha2003}
{\sc Chatha, K.~A.}
\newblock {\em Multi-Process Modelling Approach to Complex Organisation
  Design}.
\newblock PhD thesis, Loughborough University, 2003.

\bibitem{colfer2010mirroring}
{\sc Colfer, L., and Baldwin, C.~Y.}
\newblock The mirroring hypothesis: Theory, evidence and exceptions.
\newblock {\em working paper\/} (Feb. 2010).

\bibitem{Cross2005}
{\sc Cross, R., Liedtka, J., and Weiss, L.}
\newblock A practical guide to social networks.
\newblock {\em Harvard Business Review\/} (2005), --.

\bibitem{Crowston03thesocial}
{\sc Crowston, K., and Howison, J.}
\newblock The social structure of open source software development teams.
\newblock In {\em First Monday\/} (2003).

\bibitem{crowston06}
{\sc Crowston, K., and Howison, J.}
\newblock Assessing the health of open source communities.
\newblock {\em IEEE Computer 39}, 5 (2006), 89--91.

\bibitem{BoerV11}
{\sc de~Boer, R.~C., and van Vliet, H.}
\newblock Experiences with semantic wikis for architectural knowledge
  management.
\newblock In {\em WICSA\/} (2011), IEEE Computer Society, pp.~32--41.

\bibitem{DingLTV15}
{\sc Ding, W., Liang, P., Tang, A., and van Vliet, H.}
\newblock Causes of architecture changes: An empirical study through the
  communication in oss mailing lists.
\newblock In {\em SEKE\/} (2015), H.~Xu, Ed., KSI Research Inc. and Knowledge
  Systems Institute Graduate School, pp.~403--408.

\bibitem{ErikAndriessen2005}
{\sc Erik~Andriessen, J.}
\newblock Archetypes of knowledge communities.
\newblock {\em Communities and Technologies 2005\/} (2005), 191--213.

\bibitem{Faludi2006}
{\sc Faludi, A., and Watherhout, B.}
\newblock Debating evidence-based planning.
\newblock {\em disP 42}, 2 (2006), 71--72.

\bibitem{FouladiS08}
{\sc Fouladi, R.~T., and Steiger, J.~H.}
\newblock The fisher transform of the pearson product moment correlation
  coefficient and its square: Cumulants, moments, and applications.
\newblock {\em Communications in Statistics - Simulation and Computation 37}, 5
  (2008), 928--944.

\bibitem{situated}
{\sc Gallagher, S.}
\newblock Introduction: The arts and sciences of the situated body.
\newblock {\em Janus Head 9}, 2 (2006), 1--2.

\bibitem{gelman2013garden}
{\sc Gelman, A., and Loken, E.}
\newblock The garden of forking paths: Why multiple comparisons can be a
  problem, even when there is no ?fishing expedition? or ?p-hacking? and the
  research hypothesis was posited ahead of time.

\bibitem{gnatzy2011delphi}
{\sc Gnatzy, T.}
\newblock {\em Delphi studies and scenario planning: methodological
  advancements and cases}.
\newblock PhD thesis, 2011.

\bibitem{snowball}
{\sc Goodman, L.~A.}
\newblock Snowball sampling.
\newblock {\em The Annals of Mathematical Statistics 32}, 1 (1961), pp.
  148--170.

\bibitem{soa}
{\sc Group, T.~O.}
\newblock Soa source book.
\newblock {\em http://www.opengroup.org/projects/soa-book/\/}.

\bibitem{herbsleb1999beyondconwayslaw}
{\sc Herbsleb, J., and Grinter, R.}
\newblock Architectures, coordination, and distance: Conway's law and beyond.
\newblock {\em Software, IEEE 16}, 5 (sep/oct 1999), 63 --70.

\bibitem{HerbslebG99}
{\sc Herbsleb, J.~D., and Grinter, R.~E.}
\newblock Splitting the organization and integrating the code: Conway's law
  revisited.
\newblock In {\em ICSE\/} (1999), B.~W. Boehm, D.~Garlan, and J.~Kramer, Eds.,
  ACM, pp.~85--95.

\bibitem{alonetogether}
{\sc Howison, J., and University., S.}
\newblock {\em Alone together: A socio-technical theory of motivation,
  coordination and collaboration technologies in organizing for free and open
  source software development [microform]}.

\bibitem{IivariI10}
{\sc Iivari, J., and Iivari, N.}
\newblock Organizational culture and the deployment of agile methods: The
  competing values model view.
\newblock In {\em Agile Software Development}, T.~Dings�yr, T.~Dyb�, and
  N.~B. Moe, Eds. Springer, 2010, pp.~203--222.

\bibitem{Jassowski12}
{\sc Jassowski, M.}
\newblock Organizational dynamics: Understanding the impact of organizational
  structure in team productivity.
\newblock {\em IEEE Design \& Test of Computers 29}, 3 (2012), 52--59.

\bibitem{Kahneman2012}
{\sc Kahneman, D.}
\newblock {\em Thinking, Fast and Slow}.
\newblock Penguin Books, London, 2012.

\bibitem{Kazman15debt}
{\sc Kazman, R., Cai, Y., Mo, R., Feng, Q., Xiao, L., Haziyev, S., Fedak, V.,
  and Shapochka, A.}
\newblock A case study in locating the architectural roots of technical debt.
\newblock In {\em Proceedings of the International Conference on Software
  Engineering 2015\/} (2015).

\bibitem{Koch92}
{\sc Koch, G.~R.}
\newblock Software engineering as an organisational challenge.
\newblock In {\em Experimental Software Engineering Issues\/} (1992), H.~D.
  Rombach, V.~R. Basili, and R.~W. Selby, Eds., vol.~706 of {\em Lecture Notes
  in Computer Science}, Springer, pp.~62--66.

\bibitem{krippendorff04}
{\sc Krippendorff, K.}
\newblock {\em Content Analysis: An Introduction to Its Methodology (second
  edition)}.
\newblock Sage Publications, 2004.

\bibitem{KwanCD12}
{\sc Kwan, I., Cataldo, M., and Damian, D.}
\newblock Conway's law revisited: The evidence for a task-based perspective.
\newblock {\em IEEE Software 29}, 1 (2012), 90--93.

\bibitem{dsm}
{\sc LaMantia, M.~J., Cai, Y., MacCormack, A., and Rusnak, J.}
\newblock Analyzing the evolution of large-scale software systems using design
  structure matrices and design rule theory: Two exploratory cases.
\newblock In {\em WICSA\/} (2008), IEEE Computer Society, pp.~83--92.

\bibitem{LanubileCE13}
{\sc Lanubile, F., Calefato, F., and Ebert, C.}
\newblock Group awareness in global software engineering.
\newblock {\em IEEE Software 30}, 2 (2013), 18--23.

\bibitem{LanubileEPV10}
{\sc Lanubile, F., Ebert, C., Prikladnicki, R., and Vizca�no, A.}
\newblock Collaboration tools for global software engineering.
\newblock {\em IEEE Software 27}, 2 (2010), 52--55.

\bibitem{Lehman1980}
{\sc Lehman, M.~M.}
\newblock On understanding laws, evolution and conservation in the large
  program life cycle.
\newblock {\em Journal of Systems and Software 1}, 3 (1980), 213--221.

\bibitem{sambrook}
{\sc Lim, M., Griffiths, G., and Sambrook, S.}
\newblock Organizational structure for the twenty-first century.
\newblock Presented the annual meeting of The Institute for Operations Research
  and The Management Sciences., AMR, pp.~125--181.

\bibitem{cacmqualcomm}
{\sc Magnoni, S., Tamburri, D.~A., Di~Nitto, E., and Kazman, R.}
\newblock Analyzing quality models for software communities.
\newblock {\em Communications of the ACM -\/} (2017), Under Submission.

\bibitem{MaranzatoNH11}
{\sc Maranzato, R., Neubert, M., and Herculano, P.}
\newblock Moving back to scrum and scaling to scrum of scrums in less than one
  year.
\newblock In {\em OOPSLA Companion\/} (2011), C.~V. Lopes and K.~Fisher, Eds.,
  ACM, pp.~125--130.

\bibitem{MartiniBC14}
{\sc Martini, A., Bosch, J., and Chaudron, M.}
\newblock Architecture technical debt: Understanding causes and a qualitative
  model.
\newblock In {\em EUROMICRO-SEAA\/} (2014), IEEE Computer Society, pp.~85--92.

\bibitem{MatsubaraH14}
{\sc Matsubara, Y., and Hasida, K.}
\newblock K-repeating substrings: a string-algorithmic approach to
  privacy-preserving publishing of textual data.
\newblock In {\em PACLIC\/} (2014), W.~Aroonmanakun, P.~Boonkwan, and
  T.~Supnithi, Eds., The PACLIC 28 Organizing Committee and PACLIC Steering
  Committee / ACL / Department of Linguistics, Faculty of Arts, Chulalongkorn
  University, pp.~658--667.

\bibitem{sna}
{\sc Meneely, A., Williams, L., Snipes, W., and Osborne, J.~A.}
\newblock Predicting failures with developer networks and social network
  analysis.
\newblock In {\em SIGSOFT FSE\/} (2008), M.~J. Harrold and G.~C. Murphy, Eds.,
  ACM, pp.~13--23.

\bibitem{Meyer14}
{\sc Meyer, B.}
\newblock {\em Agile!}
\newblock Springer, Cham, 2014.

\bibitem{Mislove2007}
{\sc Mislove, A., Marcon, M., Gummadi, K.~P., Druschel, P., and Bhattacharjee,
  B.}
\newblock Measurement and analysis of online social networks.
\newblock In {\em Proceedings of the 7th ACM SIGCOMM conference on Internet
  measurement\/} (New York, NY, USA, 2007), IMC '07, ACM, pp.~29--42.

\bibitem{MoCKX15}
{\sc Mo, R., Cai, Y., Kazman, R., and Xiao, L.}
\newblock Hotspot patterns: The formal definition and automatic detection of
  architecture smells.
\newblock In {\em WICSA\/} (2015), L.~Bass, P.~Lago, and P.~Kruchten, Eds.,
  IEEE Computer Society, pp.~51--60.

\bibitem{Moha2010}
{\sc Moha, N., Gu\'eh\'eneuc, Y.-G., Duchien, L., and Le~Meur, A.-F.}
\newblock Decor: A method for the specification and detection of code and
  design smells.
\newblock {\em IEEE Transactions on Software Engineering 36}, 1 (Jan.-Feb.
  2010), 20--36.

\bibitem{GVK027241513}
{\sc Mohr, L.}
\newblock {\em Explaining organizational behavior}, 1. ed~ed.
\newblock The Jossey-Bass social and behavioral science series. Jossey-Bass,
  San Francisco [u.a.], 1982.

\bibitem{MoonH14}
{\sc Moon, E., and Howison, J.}
\newblock Modularity and organizational dynamics in open source software (oss)
  production.
\newblock In {\em AMCIS\/} (2014), Association for Information Systems.

\bibitem{XPImpact}
{\sc M{\"u}lller, M.}
\newblock A preliminary study on the impact of a pair design phase on pair
  programming and solo programming.

\bibitem{nagappan08}
{\sc Nagappan, N., Murphy, B., and Basili, V.}
\newblock The influence of organizational structure on software quality: an
  empirical case study.
\newblock In {\em International conference on Software engineering\/} (Leipzig,
  Germany, May 2008), IEEE, pp.~521--530.

\bibitem{NevoW05}
{\sc Nevo, D., and Wand, Y.}
\newblock Organizational memory information systems: a transactive memory
  approach.
\newblock {\em Decision Support Systems 39}, 4 (2005), 549--562.

\bibitem{rel2}
{\sc Nguyen, T., Wolf, T., and Damian, D.}
\newblock Global software development and delay: Does distance still matter?
\newblock {\em Global Software Engineering, 2008. ICGSE 2008. IEEE
  International Conference on\/} (Aug. 2008), 45--54.

\bibitem{PaasivaaraLH12}
{\sc Paasivaara, M., Lassenius, C., and Heikkil�, V.}
\newblock Inter-team coordination in large-scale globally distributed scrum: do
  scrum-of-scrums really work?
\newblock In {\em ESEM\/} (2012), P.~Runeson, M.~H�st, E.~Mendes, A.~A.
  Andrews, and R.~Harrison, Eds., ACM, pp.~235--238.

\bibitem{Polk11}
{\sc Polk, R.}
\newblock Agile and kanban in coordination.
\newblock In {\em AGILE\/} (2011), IEEE Computer Society, pp.~263--268.

\bibitem{Prandy2000}
{\sc Prandy, K.}
\newblock {\em The social interaction approach to the measurement and analysis
  of social stratification}.
\newblock No.~19. 10 2000.

\bibitem{Prikladnicki12}
{\sc Prikladnicki, R.}
\newblock Propinquity in global software engineering: examining perceived
  distance in globally distributed project teams.
\newblock {\em Journal of Software Maintenance 24}, 2 (2012), 119--137.

\bibitem{QuintanePRM13}
{\sc Quintane, E., Pattison, P., Robins, G., and Mol, J.~M.}
\newblock Short- and long-term stability in organizational networks: Temporal
  structures of project teams.
\newblock {\em Social Networks 35}, 4 (2013), 528--540.

\bibitem{RamchandP12}
{\sc Ramchand, A., and Pan, S.~L.}
\newblock The co-evolution of communities of practice and knowledge management
  in organizations.
\newblock {\em DATA BASE 43}, 1 (2012), 8--23.

\bibitem{Ratner2009}
{\sc Ratner, B.}
\newblock The correlation coefficient: Its values range between +1/−1, or do
  they?
\newblock {\em Journal of Targeting, Measurement and Analysis for Marketing
  17}, 2 (Jun 2009), 139--142.

\bibitem{valbook}
{\sc Richardson, I., Casey, V., Burton, J., and McCaffery, F.}
\newblock Global software engineering: A software process approach.
\newblock In {\em Collaborative Software Engineering}, I.~Mistrik, J.~Grundy,
  A.~van~der Hoek, and J.~Whitehead, Eds. Springer, January 2010.

\bibitem{glass}
{\sc Rost, J., and Glass, R.~L.}
\newblock The impact of subversive stakeholders on software projects.
\newblock {\em Commun. ACM 52}, 7 (July 2009), 135--138.

\bibitem{Ruikar2009}
{\sc Ruikar, K., Koskela, L., and Sexton, M.}
\newblock Communities of practice in construction case study organisations:
  Questions and insights.
\newblock {\em Construction Innovation 9}, 4 (2009), 434--.

\bibitem{Ruuska2003}
{\sc Ruuska, I., and Vartiainen, M.}
\newblock Communities and other social structures for knowledge sharing: A case
  study in an internet consultancy company.
\newblock {\em Communities and Technologies\/} (2003), 163--183.

\bibitem{Ryynanen12}
{\sc Ryyn�nen, H.}
\newblock A social network analysis of internal communication in a matrix
  organisation - the context of project business.
\newblock {\em IJBIS 11}, 3 (2012), 324--342.

\bibitem{SchwaberBeedle02}
{\sc Schwaber, K., and Beedle, M.}
\newblock {\em Agile Software Development with Scrum}.
\newblock Prentice Hall, Upper Saddle River, NJ, 2002.

\bibitem{situatedness}
{\sc Sierhuis, M., and Clancey, W.~J.}
\newblock Knowledge, practice, activities and people.
\newblock 1997.

\bibitem{Svensson2012}
{\sc Svensson, J.}
\newblock {Quality control of coding of survey responses at Statistics Sweden}.
\newblock In {\em Proceedings of the European Conference on Quality in Official
  Statistics - Q2012\/} (2012).

\bibitem{SyeedH13}
{\sc Syeed, M. M.~M., and Hammouda, I.}
\newblock Socio-technical congruence in oss projects: Exploring conway's law in
  freebsd.
\newblock In {\em OSS\/} (2013), E.~Petrinja, G.~Succi, N.~E. Ioini, and
  A.~Sillitti, Eds., vol.~404 of {\em IFIP Advances in Information and
  Communication Technology}, Springer, pp.~109--126.

\bibitem{icgseoss}
{\sc Tamburri, D.~A., di~Nitto, E., Lago, P., and van Vliet, H.}
\newblock On the nature of the {GSE} organizational social structure: an
  empirical study.
\newblock {\em proceedings of the 7th {IEEE} International Conference on Global
  Software Engineering\/} (2012), 1--10.

\bibitem{archcomm}
{\sc Tamburri, D.~A., Kazman, R., and Fahimi, H.}
\newblock The architect's role in community shepherding.
\newblock {\em IEEE Software 33}, 6 (2016), 70--79.

\bibitem{jisaspecissue}
{\sc Tamburri, D.~A., Kruchten, P., Lago, P., and van Vliet, H.}
\newblock Social debt in software engineering: insights from industry.
\newblock {\em J. Internet Services and Applications 6}, 1 (2015), 10:1--10:17.

\bibitem{specissue}
{\sc Tamburri, D.~A., Lago, P., and van Vliet, H.}
\newblock Uncovering latent social communities in software development.
\newblock {\em IEEE Software 30}, 1 (jan.-feb. 2013), 29 --36.

\bibitem{ossslr}
{\sc Tamburri, D.~A., Lago, P., and Vliet, H.~v.}
\newblock Organizational social structures for software engineering.
\newblock {\em ACM Comput. Surv. 46}, 1 (July 2013), 3:1--3:35.

\bibitem{TeradaS13}
{\sc Terada, A., and Sese, J.}
\newblock Bonferroni correction hides significant motif combinations.
\newblock In {\em BIBE\/} (2013), IEEE Computer Society, pp.~1--4.

\bibitem{TorchianoRM11}
{\sc Torchiano, M., Ricca, F., and Marchetto, A.}
\newblock Is my project's truck factor low?: theoretical and empirical
  considerations about the truck factor threshold.
\newblock In {\em WETSoM\/} (2011), G.~Concas, E.~D. Tempero, H.~Zhang, and
  M.~D. Penta, Eds., ACM, pp.~12--18.

\bibitem{vanVliet03}
{\sc van Vliet, H.}
\newblock {\em Software Engineering: principles and practice, 2nd Ed.}
\newblock John Wiley, 2000.

\bibitem{WangK01}
{\sc Wang, X., and Karten, W.}
\newblock Organizational structure triangle stability.
\newblock In {\em Human.Society@Internet\/} (2001), W.~Kim, T.~W. Ling, Y.-J.
  Lee, and S.-S. Park, Eds., vol.~2105 of {\em Lecture Notes in Computer
  Science}, Springer, pp.~299--306.

\bibitem{WeinreichG16}
{\sc Weinreich, R., and Groher, I.}
\newblock Software architecture knowledge management approaches and their
  support for knowledge management activities: A systematic literature review.
\newblock {\em Information \& Software Technology 80\/} (2016), 265--286.

\bibitem{Wenger:1998}
{\sc Wenger, E.}

\bibitem{wohlin}
{\sc Wohlin, C., Runeson, P., H\"{o}st, M., Ohlsson, M.~C., Regnell, B., and
  Wessl{\'e}n, A.}
\newblock {\em Experimentation in software engineering: an introduction}.
\newblock Kluwer Academic Publishers, Norwell, MA, USA, 2000.

\bibitem{0029933}
{\sc Wohlin, C., Runeson, P., H�st, M., Ohlsson, M.~C., and Regnell, B.}
\newblock {\em Experimentation in Software Engineering.}
\newblock Springer, 2012.

\bibitem{xiao:2016}
{\sc Xiao, L., Cai, Y., Kazman, R., Mo, R., and Feng, Q.}
\newblock Identifying and quantifying architectural debts.
\newblock In {\em 38th International Conference on Software Engineering, {ICSE}
  '16, Austin, TX - May, 2016\/} (2016).

\end{thebibliography}

\section*{Appendix: Dataset Summary Tables}
\begin{table}
\scriptsize
\centering
\caption{Outline of the sets of types (Columns 2 to 9) detected via decision-tree analysis mapped to our projects (Column 1).}\label{results}
\begin{tabular}{|c|c|c|c|c|c|c|c|c|c|}
\hline 
\textbf{project ID} & Type 1 & Type 2 & Type 3 & Type 4 & Type 5 & Type 6 & Type 7 & Type 8 & Type 9\tabularnewline
\hline 
1 & CoP  & FN  & IC  & WG  & SC  & PSC & KC &  LC & -\tabularnewline
\hline 
2 & CoP  & FN  & IC  & WG  & SC  & PSC &  SN & - & -\tabularnewline
\hline 
3 & FN & IC & PT  & SC  & PSC &  SN & - & - & -\tabularnewline
\hline 
4 & IN  & IC  & WG  & PSC  & SN & - & - & - & -\tabularnewline
\hline 
5 & FN  & NoP  & PT  & SC  & PSC  & SN & - & - & -\tabularnewline
\hline 
6 & CoP  & IN  & IC  & WG  & PSC & SN & LC & - & -\tabularnewline
\hline 
7 & CoP  & IN  & IC  & WG & PSC & SN  & LC &  SC & -\tabularnewline
\hline 
8 & CoP  & IN  & IC  & WG  & PSC  & SN  & FN & - & -\tabularnewline
\hline 
9 & IN  & IC  & WG  & PSC  & SN  & SC & - & - & -\tabularnewline
\hline 
10 & CoP  & IC  & WG  & PSC  & SN  & SC & - & - & -\tabularnewline
\hline 
11 & CoP  & IC  & WG  & PSC  & SN  & - & - & - & -\tabularnewline
\hline 
12 & CoP  & IN  & IC  & WG  & PSC  & SN &  LC & - & -\tabularnewline
\hline 
13 & IN & IC  & FN  & WG & PSC  & SC  & KC & - & -\tabularnewline
\hline 
14 & IC  & WG  & PSC & SC  & KC &  FN & - & - & -\tabularnewline
\hline 
15 & IN & IC  & FN & PSC  & SN  & LC & - & - & -\tabularnewline
\hline 
16 & - & - & - & - & - & - & - & - & -\tabularnewline
\hline 
17 & CoP  & IN & IC  & WG  & SN  & FG & - & - & -\tabularnewline
\hline 
18 & IC  & PT  & PSC  & SN & LC & - & - & - & -\tabularnewline
\hline 
19 & CoP  & IC  & WG  & PSC  & SN  & SC & - & - & -\tabularnewline
\hline 
20 & CoP  & IN  & IC  & WG  & PSC  & SN  & - & - & -\tabularnewline
\hline 
21 & CoP  & IN  & IC  & WG  & PSC  & SN  & LN & - & -\tabularnewline
\hline 
22 & FN  & IC  & WG  & PSC & LC  & SN & - & - & -\tabularnewline
\hline 
23 & CoP  & IN  & NoP & WG  & SC & PSC & KC &  LC & -\tabularnewline
\hline 
24 & CoP  & IN  & NoP & WG  & SC & PSC & KC &  LC & FN\tabularnewline
\hline 
25 & CoP  & IN & IC  & WG  & SN  & FG & - & - & -\tabularnewline
\hline 
26 & CoP  & IN & IC  & WG  & SN  & FG & - & - & -\tabularnewline
\hline 
27 & FN & IC & PT  & SC  & FG & KC & - & - & -\tabularnewline
\hline 
28 & IC  & PT  & SC & FG  & KC & - & - & - & -\tabularnewline
\hline 
20 & IN  & IC  & WG  & PSC  & KC & LC & - & - & -\tabularnewline
\hline 
30 & CoP  & IN  & NoP & PT  & PSC  & SN  & - & - & -\tabularnewline
\hline 
\end{tabular}
\end{table}


\begin{table}
\scriptsize
\centering
\caption{reported architecture issues (columns 2 and following) per project (column 1), an overview.}\label{issuesoverview}
\begin{tabular}{|>{\raggedright}p{.2cm}|>{\raggedright}p{.8cm}|>{\raggedright}p{1cm}|>{\centering}p{.85cm}|>{\centering}p{1cm}|>{\centering}p{.9cm}|>{\centering}p{.78cm}|>{\centering}p{1.2cm}|>{\centering}p{.9cm}|c|>{\centering}p{1.3cm}|}
\hline 
\emph{IDs} & \textbf{Imposs. Swap} & \textbf{Untrace. Bus. Req.} & \textbf{Sloppy Mod.} & \textbf{Unscalab. Arch.} & \textbf{God-Class} & \textbf{Spike-Centric} & \textbf{Unmod. Core} & \textbf{Quality-Implicit} & \textbf{Monolith} & \textbf{Insens. Info. Spread.}\tabularnewline
\hline 
1 & x & x & x & x & x & x & x & x & - & -\tabularnewline
\hline 
2 & x & x & x & x & - & - & - & - & - & -\tabularnewline
\hline 
3 & x & - & - & - & x & - & - & - & - & -\tabularnewline
\hline 
4 & x & x & x & - & - & x & - & - & x & x\tabularnewline
\hline 
5 & x & x & - & x & - & - & x & - & x & x\tabularnewline
\hline 
6 & x & - & - & - & x & - & - & x & - & -\tabularnewline
\hline 
7 & - & x & - & - & x & - & x & - & x & x\tabularnewline
\hline 
8 & - & x & x & x & - & - & x & - & - & x\tabularnewline
\hline 
9 & x & x & x & x & - & - & x & - & x & -\tabularnewline
\hline 
10 & x & - & x & x & - & x & x & - & x & x\tabularnewline
\hline 
11 & x & - & x & - & - & - & - & x & - & x\tabularnewline
\hline 
12 & x & - & - & - &  & - & - & - & - & x\tabularnewline
\hline 
13 & x & - & x & - & x & - & x & - & x & x\tabularnewline
\hline 
14 & x & - & - & - & - & - & x & - & - & x\tabularnewline
\hline 
15 & x & - & x & - & - & - & - & - & - & -\tabularnewline
\hline 
16 & - & x & x & - & x & - & x & - & - & -\tabularnewline
\hline 
17 & x & - & x & - & - & - & x & - & - & -\tabularnewline
\hline 
18 & x & - & x & - & - & - & x & - & - & x\tabularnewline
\hline 
19 & - & - & x & - & - & - & x & - & - & -\tabularnewline
\hline 
20 & - & - & x & - & - & - & x & - & - & -\tabularnewline
\hline 
21 & - & x & - & - & - & - & - & - & - & -\tabularnewline
\hline 
22 & - & x & - & - & x & - & x & - & - & -\tabularnewline
\hline 
23 & - & x & - & - & - & - & x & - & - & -\tabularnewline
\hline 
24 & - & x & - & - & - & - & x & - & - & -\tabularnewline
\hline 
25 & x & x & - & - & x & - & - & - & - & -\tabularnewline
\hline 
26 & x & - & - & - & x & - & x & - & - & -\tabularnewline
\hline 
27 & x & x & - & - & x & - & - & - & - & -\tabularnewline
\hline 
28 & x & - & - & - & x & - & x & - & - & x\tabularnewline
\hline 
29 & x & x & - & - & x & - & - & - & - & x\tabularnewline
\hline 
30 & x & x & x & x & x & x & x & x & x & x\tabularnewline
\hline 
\end{tabular}
\end{table}

\end{document}